\newif\ifconfver
\confverfalse      

\newif\ifplainver  
\plainvertrue

\newif\ifhide  
\hidetrue

\ifplainver
    \confverfalse   
\fi

\ifconfver
     \documentclass[10pt,twocolumn,twoside]{IEEEtran}
\else
    \ifplainver
        \documentclass[11pt]{article}
        \usepackage{fullpage}
    \else
        \documentclass[12pt,draftcls,onecolumn]{IEEEtran}
    \fi
\fi

\usepackage{calc,amsfonts,amssymb,amsmath,bm,url,color,theorem,graphicx,cite}
\usepackage{psfrag,subfigure,float}
\usepackage{algorithm,soul}
\usepackage{algorithmicx}
\usepackage{algpseudocode}
\usepackage{soul}
\usepackage{enumerate}
\usepackage{multirow,multicol}
\usepackage{marvosym}
\usepackage{nidanfloat}

\definecolor{orange}{RGB}{255,107,0}
\def\blue{\color{blue}}
\def\red{\color{red}}

\def\brown{\color{brown}}

\usepackage{xcolor,cite,etoolbox}
\makeatletter
\pretocmd\@bibitem{\color{black}\csname keycolor#1\endcsname}{}{\fail}
\newcommand\citecolor[1]{\@namedef{keycolor#1}{\color{blue}}}
\makeatother

\newtheorem{Fact}{Fact}

\theorembodyfont{\rmfamily}

\newcommand\bW{\ensuremath{{\bm W}}}

\newcommand\bx{\ensuremath{{\bm x}}}
\newcommand\by{\ensuremath{{\bm y}}}
\newcommand\bG{\ensuremath{{\bm G}}}

\newcommand\bX{\ensuremath{{\bm X}}}
\newcommand\bZ{\ensuremath{{\bm Z}}}

\newcommand\ba{\ensuremath{{\bm a}}}

\newcommand\bA{\ensuremath{{\bm A}}}

\newcommand\bB{\ensuremath{{\bm B}}}

\newcommand\bF{\ensuremath{{\bm F}}}

\newcommand\bYh{\ensuremath{{\bm Y}_{\rm H}}}
\newcommand\bYm{\ensuremath{{\bm Y}_{\rm M}}}

\newcommand\bym{\ensuremath{{\bm y}_{\rm M}}}

\newcommand\bVh{\ensuremath{{\bm V}_{\rm H}}}
\newcommand\bVm{\ensuremath{{\bm V}_{\rm M}}}
\newcommand\Lh{\ensuremath{L_{\rm H}}}
\newcommand\Mm{\ensuremath{M_{\rm M}}}

\newcommand\bY{\ensuremath{{\bm Y}}}

\newcommand\bU{\ensuremath{{\bm U}}}
\newcommand\bs{\ensuremath{{\bm s}}}
\newcommand\bS{\ensuremath{{\bm S}}}

\newcommand{\Rbb}{\mathbb{R}}

\newcommand{\Sbb}{\mathbb{S}}

\newcommand{\setX}{\mathcal{X}}

\newcommand{\Diag}{\mathrm{Diag}}

\newcommand\bLam{\ensuremath{{\bm \Lambda}}}

\newcommand{\bzero}{{\bm 0}}
\newcommand{\bone}{{\bm 1}}
\newcommand{\bI}{{\bm I}}

\newcommand\bigO{\ensuremath{{\mathcal{O}}}}

\newcommand{\lammax}{\lambda_{\rm max}}

\usepackage{tikz}
\usetikzlibrary{arrows.meta}
\usetikzlibrary{decorations}
\usetikzlibrary{calc}
\usetikzlibrary{shapes.geometric}
\usetikzlibrary{external}

\begin{document}

\bibliographystyle{IEEEtran}

\newcommand{\papertitle}{
Hyperspectral Super-Resolution via
Global-Local Low-Rank Matrix Estimation
}

\newcommand{\paperabstract}{
Hyperspectral super-resolution (HSR) is a problem that aims to estimate an image of high spectral and spatial resolutions from a pair of co-registered  multispectral (MS) and hyperspectral (HS) images, which have coarser spectral and spatial resolutions, respectively.
In this paper we pursue a low-rank matrix estimation approach for HSR.
We assume that the spectral-spatial matrices associated with the whole image and the local areas of the image have low-rank structures.
The local low-rank assumption, in particular, has the aim of providing a more flexible model for accounting for local variation effects due to endmember variability.
We formulate the HSR problem as a global-local rank-regularized least-squares problem.
By leveraging on the recent advances in non-convex large-scale optimization, namely, the smooth Schatten-$p$ approximation and the accelerated majorization-minimization method,
we develop an efficient algorithm for the global-local low-rank problem.
Numerical experiments on synthetic, semi-real and real data show that the proposed algorithm outperforms a number of benchmark algorithms in terms of recovery performance.
}


\ifplainver


    \title{\papertitle}

    \author{
    Ruiyuan Wu$^\dag$, Wing-Kin Ma$^\dag$, Xiao Fu$^\ddag$, and Qiang Li$^\star$
    \\ ~ \\
    $^\dag$Department of Electronic Engineering, The Chinese University of Hong Kong, \\
    Hong Kong SAR of China \\ ~ \\
    $^\ddag$School of Electrical Engineering and Computer Science, \\
    Oregon State University, Corvallis, USA \\  ~ \\
    $^\star$School of Information and Communications Engineering, \\
    University of Electronic Science and Technology of China, China \\ ~ \\
    E-mails: rywu@ee.cuhk.edu.hk, wkma@ee.cuhk.edu.hk, \\
    xiao.fu@oregonstate.edu,
    lq@uestc.edu.cn  \\
    }

    \maketitle

    \begin{abstract}
    \paperabstract
    \end{abstract}

\else
    \title{\papertitle}

    \ifconfver \else {\linespread{1.1} \rm \fi

    \author{Ruiyuan Wu, Wing-Kin Ma, Xiao Fu, and Qiang Li
    	\thanks{
    		This work was supported by Project \#MMT-8115059 of the Shun Hing
    		Institute of Advanced Engineering, The Chinese University of Hong Kong.
    		Xiao Fu was supported in part by the National Science Foundation under projects NSF ECCS 1608961 and NSF ECCS 1808159, and in part by the Army Research Office under project ARO W911NF-19-1-0247.
    	}
    	\thanks{
    		Ruiyuan Wu is with
    		Department of Electronic Engineering, The Chinese University of Hong
    		Kong, Shatin, N.T., Hong Kong SAR of China.
    		E-mail: rywu@link.cuhk.edu.hk, Tel: +852-31638271, Fax: +852-26035558.
    	}
    	\thanks{
    	Wing-Kin Ma is the corresponding author. Address:
    	Department of Electronic Engineering, The Chinese University of Hong
    	Kong, Shatin, N.T., Hong Kong SAR of China.
    	E-mail: wkma@ieee.org, Tel: +852-31634350, Fax: +852-26035558.
    	}
    	\thanks{
    		Xiao Fu is with School of Electrical Engineering and Computer Science,
    		Oregon State University, Corvallis, USA.
    		E-mail: xiao.fu@oregonstate.edu,  Tel: +1 (541) 737-3617, Fax: +1 (541) 737-1300.
    	}
    	\thanks{
    		Qiang Li is with School of Information and Communications Engineering,
            University of Electronic Science and Technology of China, Chengdu,
            China, and also with Peng Cheng Laboratory, Shenzhen, China. E-mail:
            lq@uestc.edu.cn, Tel: +86-28-61830156, Fax: +86-28-61831665.
		}
	}

    \maketitle

    \ifconfver \else
        \begin{center} \vspace*{-2\baselineskip}
        \end{center}
    \fi

    \begin{abstract}
    \paperabstract
    \\\\
    \end{abstract}

    \vspace*{-2em}


    \ifconfver \else \IEEEpeerreviewmaketitle} \fi

 \fi

\ifconfver \else
    \ifplainver \else
        \newpage
\fi \fi


\section{Introduction}

Hyperspectral (HS) sensors have limited spatial resolution as a tradeoff for achieving high spectral resolution.
Such a tradeoff is due to hardware limitations and the measurement mechanism.
In a nutshell, a certain amount of light energy reflected by the scene is required for each spectral band of an HS pixel to achieve sufficiently high SNRs.
For a sensor with coarse spectral resolution, enough energy can be acquired from a small area by accumulating energy of a wide range of spectral bands.
When the spectral resolution increases, the area sensed by a pixel needs to be enlarged to acquire the same amount of energy for each spectral band, which leads to a lower spatial resolution.
How to enhance the spatial resolution of HS images has been a subject of great interest.
Recently, the idea of using an additional multispectral (MS) image---which has finer spatial resolution than the HS but possesses only several coarse spectral bands---for HS spatial resolution enhancement has shed new light on the subject.
This MS-aided enhancement problem is called {\em hyperspectral super-resolution} (HSR) or HS-MS data fusion.
One approach for HSR is to adapt pansharpening techniques in fusion of panchromatic and HS images \cite{loncan2015hyperspectral}.
Another approach, which is currently more popular, is based on low-dimensional data models.
The low-dimensional model approach assumes that the spectral pixels of the target high-spatial-resolution HS image,
or the super-resolution (SR) image for short,
lie in a low-dimensional subspace.
This assumption
is particularly reasonable
in the linear spectral mixture scenario:
Since the aforesaid scenario has every spectral pixel described as a linear combination of the spectral signatures of the underlying endmembers,
the spectral pixels lie in a subspace spanned by the endmember spectral signatures.
Also, since the number of endmembers is often much smaller than the number of HS spectral bands,
the subspace dimension is low.
The low-dimensional model may also be applicable to some classes of nonlinear spectral mixtures such as the bilinear mixture model \cite{dobigeon2013nonlinear,heylen2014review}.
The low-dimensional model approach has strong connections to hyperspectral unmixing (HU).
To be specific, insights and methods in HU are quite often used in the low-dimensional model approach.
A comparative review has shown that the low-dimensional model approach can lead to better recovery than those from the pansharpening approach, assuming no or negligible HS-MS co-registration error \cite{loncan2015hyperspectral}.

To facilitate our discussion later,
in this paper we taxonomize the existing low-dimensional model-based HSR methods into two types.
\begin{enumerate}[1.]
\item {\em Matrix factorization:} \
This type of methods models the spectral-spatial matrix of the SR image as a product of two matrix factors---one being the spectral dictionary, and another the coefficients for low-dimensional representation, and it seeks to jointly estimate the two matrix factors from the observed HS-MS pair.
Coupled non-negative matrix factorization (CNMF)~\cite{yokoya2012coupled}, a pioneering HSR method, falls into this type.
As its name suggests, CNMF exploits the non-negativity of the matrix factors. Subsequent research follows the same route and exploits other problem structures---sparsity~\cite{wycoff2013non}, the sum-to-one abundance condition from the linear spectral mixture model~\cite{wei2016multiband,wu2019hybrid}, {non-local pixel similarity~\cite{dian2019multispectral}}, and many more---to attempt to improve recovery quality.



\item {\em Dictionary-based regression:} \
This type of methods also
assumes that the spectral-spatial matrix of the SR image is the product of a spectral dictionary and the associated coefficient matrix.
The difference is that it does not seek to jointly estimate the spectral dictionary and the coefficients.
It first determines the spectral dictionary via some easy way,
and then uses that spectral dictionary to perform regression to recover the coefficients.
A typical example is HySure \cite{simoes2015convex}, which
extracts the spectral dictionary by applying vertex component analysis (VCA) \cite{nascimento2005vertex} to the HS image, and then
recovers the coefficients by applying spatial total variation-regularized linear regression to the HS-MS image pair.
Other methods include \cite{veganzones2016hyperspectral,dian2018hyperspectral}, which exploit the local low-rank structure; this will be further discussed later.
The dictionary-based regression methods are easy to implement compared to the matrix factorization methods.

\end{enumerate}
Research on these two types of methods is mostly focused on the practical aspects, and it is worthwhile to note that some specific methods have recently been shown to exhibit theoretical recovery guarantees as well \cite{hsr_recovery_ssp2018,liu2019hsr_recovery}---which supports the soundness of the low-dimensional model approach via a theoretical lens.

Matrix factorization and dictionary-based regression are considered most representative in HSR methods, although there are others.
For example,
tensor factorization has recently been studied for HSR \cite{li2018fusing,kanatsoulis2018hyperspectral,prevost2019coupled,dian2019learning,dian2019hyperspectral,dian2019nonlocal}.
The tensor model is also a low-dimensional model, and it exploits not only the spectral-spatial structure but also the two-dimensional spatial structure.
Tensor factorization is shown to exhibit favorable sufficient conditions on exact recovery guarantees \cite{kanatsoulis2018hyperspectral,prevost2019coupled}.
In addition, deep learning for HSR has most recently gained growing interest \cite{palsson2017multispectral,lanaras2018super,dian2018deep}.


Under the low-dimensional model, HSR can be seen as a problem of recovering a low-rank matrix from incomplete observations;
this will be elaborated upon in Sections~\ref{sec:3} and~\ref{sec:4}.
From this perspective, the problem is nearly the same as the matrix completion problem which has drawn widespread interest in fields such as recommender systems, machine learning and mathematical optimization \cite{recht2010guaranteed,sun2016guaranteed,chen2018harnessing}.
The problem in matrix completion is that we have a matrix with many missing entries, and we aim to recover the missing entries from the available entries.
The main assumption in matrix completion is that the matrix to be recovered has low rank structure.
This assumption is the same as the low-dimensional model assumption in HSR.
In matrix completion we see two main types of methods.
One is matrix factorization, which shares the same rationale as matrix factorization for HSR.
Another is low-rank matrix estimation.
This approach does not pre-determine the target dimension of the low-dimensional subspace, or the target rank of the matrix to be recovered, which is the case in matrix factorization.
Instead of using the matrix factorization model, it seeks the minimum rank matrix for accomplishing the task.
A well-known method in low-rank matrix estimation is the nuclear norm minimization method \cite{recht2010guaranteed}.
It is a convex solution, and the idea is to approximate the hard-to-handle rank function by the nuclear norm which is convex.
Non-convex rank approximation, such as the Schatten-$p$ approximation, was also considered for approximating the low-rank problem better \cite{mohan2012iterative}.
Back to HSR, while we have seen numerous studies on matrix factorization and dictionary-based regression,
we see far less on
low-rank matrix estimation.

In addition, and as a different issue, the existing low-dimensional model-based HSR methods are usually not designed to account for the endmember variability (EV) effects due, for instance, to illumination conditions and intrinsic spectral variability of the materials \cite{zare2013endmember}.
In low-dimensional models, EV means that the spectral dictionary can vary in space.
Taking a step back to HU, we have seen studies that use the matrix factorization method to deal with endmember variability \cite{halimi2015unsupervised,thouvenin2015hyperspectral,drumetz2019spectral}.
In that regard, a possibility one can consider is to adapt such matrix factorization methods to the HSR application.
We are however unaware of such development as of the writing of this paper.

In this paper, our objective is to explore the potential of low-rank matrix estimation in HSR.
Our study is not a direct application of the existing low-rank matrix estimation methods, such as the nuclear norm minimization method.
Our formulation takes the possibility of EV into consideration.
We posit a global-local low-rank structure with the SR image,
in which not only the spectral-spatial matrix of the whole SR image has low-rank structure,
but that of each local area also has another low-rank structure.
This assumption means that each local area can have its low-dimensional representation.
The local low-rank assumption provides the model with the flexibility to account for EV.
Moreover, since the low-dimensional subspaces of the local areas should be related, particularly the neighboring ones, we also assume that the whole spectral-spatial matrix has low rank and utilize it to tie the local subspaces together.
As we will see, our global-local low-rank matrix estimation leads to a fairly clean formulation.
In comparison, if one applies the EV-present matrix factorization methods in HU to HSR, the resulting formulation would be more complicated.

The arising challenge and our proposed solution should be described.
The global-local low-rank matrix estimation problem is a non-convex large-scale problem.
For example, to recover an SR image of $100$ spectral bands and of size $200 \times 200$, our problem requires us to handle $100 \times 200^2 = 4,000,000$ optimization variables.
An efficient optimization strategy is clearly needed.
We attack the problem by leveraging on the recent advances in non-convex large-scale optimization,
namely, the smooth Schatten-$p$ approximation \cite{mohan2012iterative} and an accelerated version of the majorization-minimization (MM) method \cite{shao2018framework}.
As mentioned, the smooth Schatten-$p$ approximation, albeit non-convex, approximates rank better in comparison with the convex nuclear norm.
Also, its smooth nature enables us to access powerful machinery in smooth optimization.
The accelerated MM method is a combination of inexact MM and the accelerated projected gradient (PG) method.
Our recent research in another context \cite{shao2018framework}---which also deals with large problem sizes---has suggested that this type of accelerated methods runs very fast in practice.
Using the above two techniques,
we develop a fast algorithm called Global-Local lOw-Rank promotIng Algorithm (GLORIA).
As will be shown by numerical results, GLORIA gives competitive recovery performance compared to the state of the arts and related methods.
We conducted semi-real experiments on five different datasets, and GLORIA consistently ranks first or second in performance indicators such as peak SNR and spectral angle mappers.
We also provide results on synthetic and real data experiments, in which GLORIA also exhibits promising performance.

Before we proceed to the description of our method, we should mention related works.
First, the dictionary-based regression method in \cite{veganzones2016hyperspectral}, HSR-LDL-EIA, also utilizes some kind of local low-rank structures.
HSR-LDL-EIA considers the linear spectral mixture model and assumes that the number of endmembers in each local area is very small and no greater than the number of MS bands.
With that assumption, the HSR problem can be easily solved in a local-area-by-local-area fashion---which is what HSR-LDL-EIA does.
The local low-rank assumption used in HSR-LDL-EIA is not the same as the one used by us.
As discussed earlier, our local low-rank assumption is to cater for EV.
Second, the recent work in \cite{dian2018hyperspectral} takes insight from the local low-rank assumption in HSR-LDL-EIA and proposes a dictionary-based regression method using local nuclear norm-regularized linear regression.
Again, the local low-rank assumption in \cite{dian2018hyperspectral} is founded on the argument that the number of endmembers in each local area is small.
We should also mention the works \cite{zhou2017hyperspectral,qu2018hyperspectral} which follow similar rationales as those in \cite{veganzones2016hyperspectral,dian2018hyperspectral}.

{Let us summarize our contributions.
\begin{enumerate}[1.]
    \item
    We consider low-rank matrix estimation for HSR, which has not been studied in prior works.
    We propose a global-local low-rank approach which aims to account for EV effects.
    \item
    The global-local low-rank approach requires us to tackle a large-scale non-convex optimization problem.
    We custom-develop an efficient algorithm for such purpose, using recent advances in large-scale non-convex optimization.
    As will be shown in numerical experiments, our algorithm has competitive recovery performance.
\end{enumerate}
Readers who are interested in trying our algorithm can find the source codes at \url{https://github.com/REIYANG/GLORIA}.}

\section{Background}
\subsection{The Measurement Model}

\begin{figure}[!h]
    \centering
    \includegraphics[width=.65\linewidth]{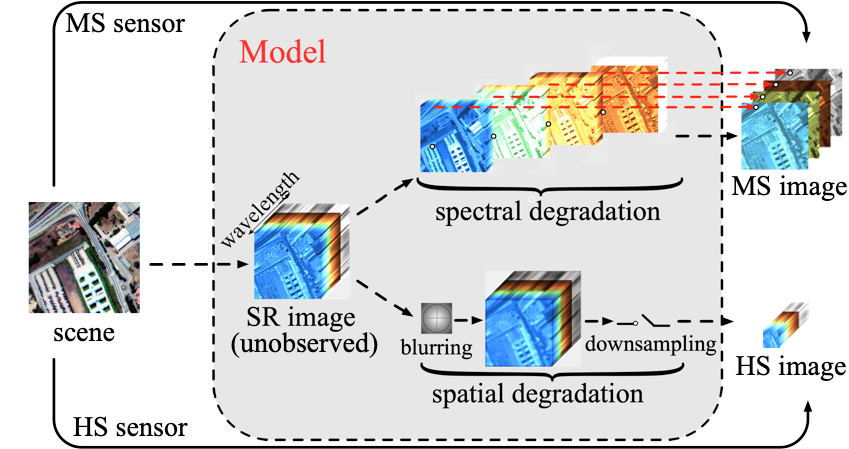}
    \caption{The HSR scenario.}
    \label{fig:model}
\end{figure}

Let us begin by providing a concise review of the background.
Fig.~\ref{fig:model} depicts the scenario.
We have a scene observed by an HS sensor and an MS sensor.
The MS sensor has a lower spectral resolution than the HS sensor,
while the HS sensor has a lower spatial resolution than the MS sensor.
The goal of HSR is to use the observed MS and HS images to construct a higher resolution image whose spectral resolution is identical to that of the HS sensor, and spatial resolution the MS.
For convenience, the image we seek to construct will be called the super-resolution (SR) image.
As a common assumption (see, e.g., \cite{yokoya2012coupled}),
the HS image is modeled as a spatially degraded version of the SR image by means of spatial blurring and down-sampling.
Also, the MS image is modeled as spectrally degraded version of the SR image by means of spectral bandpass averaging.

The HSR data model is as follows.
Assuming that the HS and MS images are co-registered, we model the HS and MS images as
\begin{subequations} \label{eq:sig_mod}
	\begin{align}
	\bYm & = \bF \bX + \bVm, \label{eq:sig_mod_m} \\
	\bYh & = \bX \bG + \bVh, \label{eq:sig_mod_h}
	\end{align}
\end{subequations}
where $\bYm \in \Rbb^{\Mm \times L}$ and $\bYh \in \Rbb^{M \times \Lh}$ are the spectral-spatial matrices of the observed  MS and HS images, respectively (resp.);
$\Mm$ and $M$ are the numbers of spectral bands of the MS and HS images, resp., with $\Mm < M$;
$L$ and $\Lh$ are the numbers of pixels of the MS and HS images, resp., with $\Lh < L$;
$\bX \in \Rbb^{M \times L}$ is the spectral-spatial matrix of the SR image;
$\bF \in \Rbb^{\Mm \times M}$ and $\bG \in \Rbb^{L \times \Lh}$ describe the measurement responses that lead to the MS and HS images, resp.;
$\bVm$ and $\bVh$ are noise.
Note that $\bF$ designates the relative spectral bandpass responses from the SR image to the MS image,
while $\bG$ specifies the spatial blurring and down-sampling responses that result in the HS image.
The measurement response matrices $\bF$ and $\bG$ are assumed to be known,
and in practice $\bF$ and $\bG$ can be acquired either by calibration \cite{yokoya2012coupled} or by estimation from the HS and MS images \cite{yokoya2013cross,simoes2015convex}.
Furthermore, the MS and HS images are measured by means of reflectance, with values lying between $0$ and $1$.
As such, we may assume that
$x_{ij} \in [0,1]$, for all $i,j$.

\subsection{The Matrix Factorization Model}

Next, we describe the matrix factorization model which is the core assumption for matrix factorization and dictionary-based regression methods in HSR.
In the matrix factorization model we assume that the SR image $\bX$ can be factored as
\begin{equation} \label{eq:mat_fac}
\bX = \bA \bS,
\end{equation}
where $\bA \in \Rbb^{M \times N}$ is the spectral dictionary;
$\bS \in \Rbb^{N \times L}$ is the coefficient matrix;
$N$ is the target rank or model order, which is pre-fixed and is often chosen to be much less than $M$ and $L$.
In many existing studies, the model \eqref{eq:mat_fac} is seen as the linear spectral mixture model in which the columns $\ba_i$'s of $\bA$ are interpreted as spectral signatures of the endmembers of the scene, and the columns $\bs_i$'s of $\bS$ the associated abundances of the pixels.
Under the model \eqref{eq:mat_fac}, the matrix factorization methods seek to find $(\bA,\bS)$ from $(\bYm,\bYh)$ by minimizing a data-fitting loss function, and thereby reconstruct $\bX$ by $\bX = \bA\bS$.
For example, in CNMF \cite{yokoya2012coupled}, the idea is to solve
\begin{equation} \label{eq:prob_cnmf}
\min_{\bA \geq \bzero, \bS \geq \bzero }  ~ \| \bYm - \bF \bA
\bS  \|_F^2 + \| \bYh - \bA \bS \bG \|_F^2,
\end{equation}
where $\bX \geq \bzero$ means that $\bX$ is element-wise non-negative; $\| \cdot \|_F$ denotes the Frobenius norm.
CNMF, as well as other matrix factorization formulations, considers structured factors to better exploit the underlying problem structure.
CNMF utilizes the fact that, under the linear spectral mixture model, $\bA$ and $\bS$ are non-negative.
Moreover, in the dictionary-based regression methods, one first determines $\bA$ from $\bYh$ by methods such as principal component analysis (PCA) or VCA.
Then, $\bS$ is estimated by solving the data-fitting problem like the one in \eqref{eq:prob_cnmf}, but with $\bA$ fixed.
In estimating $\bS$, regularization such as total variation may be added
for problem structure exploitation
\cite{simoes2015convex}.
A common, and key, concept behind the various matrix factorization and dictionary-based regression methods is that although $\bX$ is a high-dimensional matrix, we hold the belief that every spectral pixel $\bx_i$ of $\bX$ should lie in a low-dimensional subspace spanned by $\ba_1,\ldots,\ba_N$.

As an alternative view, matrix factorization and dictionary-based regression may be regarded as methods for estimating a low-rank matrix $\bX$ from the incomplete HS-MS observations.
Specifically, the low-dimensional subspace assumption with $\bX$ implies ${\rm rank}(\bX)={\rm rank}(\bA\bS)\leq N$.
\section{Global-Local Low-Rank Formulation}
\label{sec:3}

This section describes the main development of this paper, global-local low-rank matrix estimation.

\subsection{A Brief Review of Low-Rank Matrix Estimation}

The low-rank matrix estimation methods have recently become popular in the context of  matrix completion  \cite{recht2010guaranteed,chen2018harnessing}.
Let us first describe how the {\em de facto} standard in low-rank matrix estimation, namely, the nuclear norm approximation, can be applied to HSR.
In low-rank matrix estimation, we assume $\bX$ to be a low-rank matrix.
This assumption can be interpreted as requiring the columns $\bx_1,\ldots,\bx_L$ to lie in a low-dimensional subspace.
The matrix factorization model \eqref{eq:mat_fac} can also be seen as constraining $\bx_1,\ldots,\bx_L$ to lie in a low-dimensional subspace, with the subspace dimension no greater than $N$.
Hence, both the low-rank matrix estimation and matrix factorization methods exploit low-dimensional data structures.
The idea with low-rank matrix estimation is to find a low-rank $\bX$ whose data fitting loss is small.
A common low-rank matrix estimation formulation is as follows
\begin{equation} \label{eq:prob_rank_reg}
\min_{\bX \in \Rbb^{M \times L} }  ~ \ell(\bX) + \gamma  \, {\rm rank}(\bX),
\end{equation}
where $\gamma > 0$ is given and is called a regularization parameter;
$\ell: \Rbb^{M \times L} \rightarrow \Rbb$ denotes the data-fitting loss function and is given by
\begin{equation} \label{eq:loss_fn}
\ell(\bX) = \tfrac{1}{2} \| \bYm - \bF \bX \|_F^2 + \tfrac{1}{2} \| \bYh - \bX \bG \|_F^2.
\end{equation}


Let us give a brief comparison of low-rank matrix estimation and matrix factorization.
Matrix factorization methods, such as the one in \eqref{eq:prob_cnmf}, require one to pre-determine the target rank $N$.
There is no such rank constraint in low-rank matrix estimation.
The rank of $\bX$ itself serves as the regularization for the low-rank matrix recovery endeavor, and the parameter $\gamma$ controls the balance between low rankness and goodness of data fitting.

The challenge with solving problem~\eqref{eq:prob_rank_reg} is that the rank function in \eqref{eq:prob_rank_reg} is hard to handle; it is non-convex and non-differentiable.
The state of the art handles this issue by applying the nuclear norm approximation.
The reader is referred to the literature \cite{chen2018harnessing} for detailed description of the concept, and here we concisely explain the idea.
The rank of $\bX$ is identical to the number of nonzero singular values of $\bX$, and hence we can express ${\rm rank}(\bX)$ as
\begin{equation}
\label{eq:rank_definition}
    {\rm rank}(\bX)=\sum_{i=1}^{\min\{M,L\}}u(\sigma_i(\bX)),
\end{equation}
where $\sigma_i(\bX)\geq0$ denotes the $i$th largest singular value of $\bX$;
$u(y)=1$ if $y>0$, and $u(y)=0$ if $y=0$.
The idea with nuclear norm approximation is to approximate ${\rm rank}(\bX)$ by removing $u$ from~\eqref{eq:rank_definition}, which leads to the following approximate function:
\[
    \|\bX\|_*=\sum_{i=1}^{\min\{M,L\}}\sigma_i(\bX).
\]
The above function is called the nuclear norm and is known to be convex~\cite{recht2010guaranteed,chen2018harnessing}.
Applying this nuclear norm approximation of rank to problem~\eqref{eq:prob_rank_reg} gives rise to the following problem:
\begin{equation} \label{eq:prob_mn_reg}
\min_{\bX \in \Rbb^{M \times L} }  ~ \ell(\bX) + \gamma   \| \bX \|_*.
\end{equation}
The advantage of the above approximation is that it is a convex problem.
Moreover, problem~\eqref{eq:prob_mn_reg} can be efficiently solved by methods such as the accelerated proximal gradient method~\cite{toh2010accelerated} and ADMM~\cite{lin2010augmented}.

\subsection{The Global-Local Low-Rank Model}

\begin{figure}[!h]
	\centering
	\includegraphics[width=.5\linewidth]{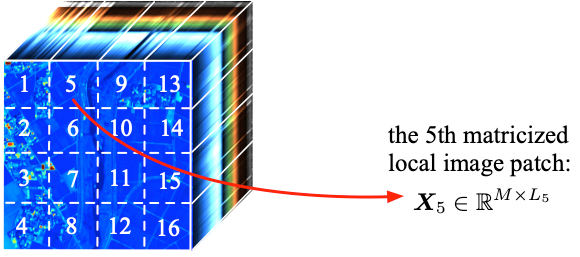}
	\caption{Segmentation of an image into patches.}
	\label{fig:patches}
\end{figure}

We consider a {\em global-local} low-rank assumption for the HSR problem.
To explain the idea, consider the illustration in Fig.~\ref{fig:patches}.
We segment the SR image into a number of local patches.
Our belief is that each local patch exhibits its own local rank structure.
Or, the low-dimensional subspace of one patch does not need to be the same as that of another.
This assumption appears to make sense since real images may have local variation effects due to EV.
Moreover, we still keep the old low-rank assumption with $\bX$.
This is because the low-dimensional subspace of one patch should be related to those of its neighboring patches, and such correlations may result in a low-rank $\bX$ in the global sense (with a higher global rank than the local ranks).
Alternatively speaking, the low-dimensional subspace of the whole $\bX$ plays the role of tying together the low-dimensional subspaces of the local patches.

To write down the global-local low-rank assumption,
we assume that the pixel indices of the image are arranged such that $\bX$ can be conveniently expressed as
\[
\bX = \begin{bmatrix}
\bX_1 & \bX_2 & \cdots & \bX_P
\end{bmatrix},
\]
where each $\bX_i \in \Rbb^{M \times L_i}$ is the spectral-spatial matrix of a local area, or local patch, of the image; $P$ is the number of patches; $L_i$ is the number of pixels of patch $i$.
For example, as illustrated in Fig.~\ref{fig:patches}, we can divide the image into equal-space rectangular blocks as our local patches.
Other ways to form the local patches, e.g., via segmentation \cite{veganzones2016hyperspectral}, may also be considered.
We assume that every $\bX_i$ is a low-rank matrix, and $\bX$ is also a low-rank matrix.

The global-local low-rank assumption stated above is fairly general and does not restrict itself to specific mixture models such as the linear spectral mixture model.
On the other hand, we can better understand the assumption by a more concrete example, in which the linear spectral mixture model is used, as follows.
Suppose we model each $\bX_i$ to follow the linear spectral mixture model
\begin{equation} \label{eq:i_am_tired}
\bX_i = \bA_i \bS_i,
\end{equation}
where $\bA_i \in \Rbb^{M \times N}$ and $\bS_i \in \Rbb^{N \times L_i}$ are the endmember and abundance matrices of patch $i$;
$N$ is the total number of endmembers in the whole SR image $\bX$.
In this model, we assume the presence of EV by allowing the endmember matrix $\bA_i$ to be different for each patch.
Note that our model assumes that EV appears at the patch level, not at the pixel level,
and this can be justified if the local region of each patch is small enough such that the endmember spectral signatures experience little or no variation within the patch.
We also want to impose an assumption that $\bA_1,\ldots,\bA_P$ are correlated, since, in reality, they should be variations of one another.
Such correlation would mean that $\bA_1,\ldots,\bA_P$ can be linearly represented by a ``global'' basis $\bB$, whose dimension is no less than $N$, but not significantly greater than $N$ owing to the correlations.
This further means that $\bX$ lives in a low-dimensional subspace with $\bB$ as its basis.
Our global low-rank assumption is to exploit the global low-dimensional structure.


Additionally, in the above motivating model example, at first sight
one would be tempted to say that the rank of $\bX_i$ is universally given by ${\rm rank}(\bX_i) = N$ (under the slightly technical premise that $\bA_i$ and $\bS_i$ have full column rank and full row rank, resp.).
In fact, it is reasonable to assume non-identical ${\rm rank}(\bX_i)$.
In practice, it is likely that each local region is composed of a small number of endmembers, rather than all of the endmembers.
Thus we can assume that among all the rows $\bs_i^1, \ldots, \bs_i^{N}$ of $\bS_i$, only $N_i$ of them are nonzero (or active).
Consequently we have ${\rm rank}(\bX_i) = N_i$
(again, under the technical premise that $\bA_i$ has full column rank and that the nonzero $\bs_i^j$'s are linearly independent),
which is our local low-rank assumption.

\begin{figure}[!hbt]
\centering
    \begin{minipage}[b]{.24\linewidth}
        \centering
        \includegraphics[width=\linewidth]{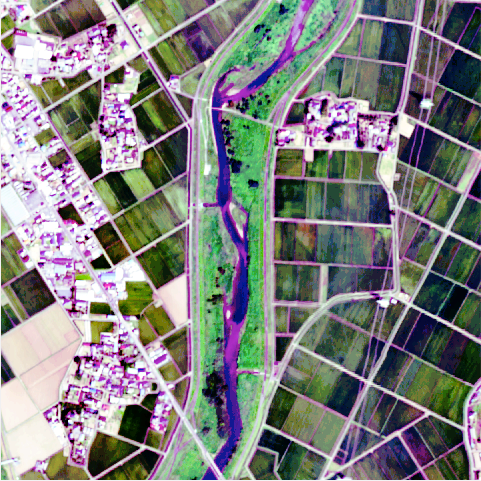}
        (a) Chikusei
    \end{minipage}
    \begin{minipage}[b]{.24\linewidth}
        \centering
        \includegraphics[width=\linewidth]{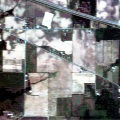}
        (b) Indian Pines
    \end{minipage}
    \begin{minipage}[b]{.24\linewidth}
        \centering
        \includegraphics[width=\linewidth]{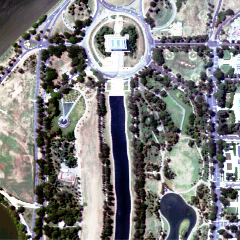}
        (c)Washington~DC~Mall
    \end{minipage}
    \begin{minipage}[b]{.24\linewidth}
        \centering
        \includegraphics[width=\linewidth]{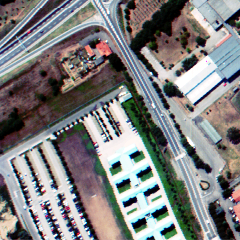}
        (d)~University of Pavia
    \end{minipage} \newline
    \begin{minipage}[b]{.24\linewidth}
        \centering
        \includegraphics[width=\linewidth]{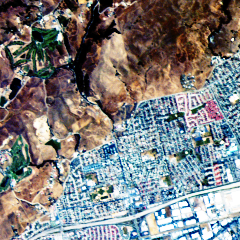}
        (e) Moffett Field
    \end{minipage}
    \begin{minipage}[b]{.24\linewidth}
        \centering
        \includegraphics[width=\linewidth]{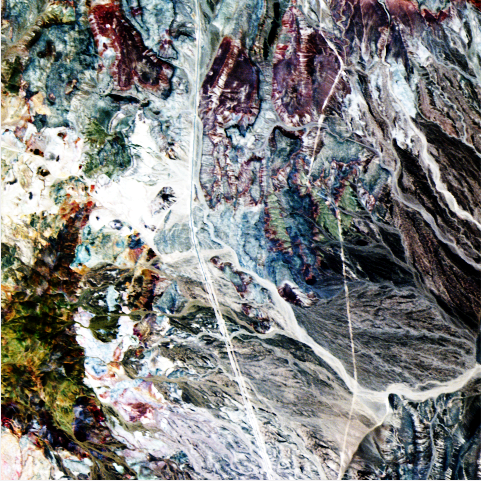}
        (f) Cuprite
    \end{minipage}\hfill~
    \caption{Color composite images of the tested HS images.}
    \label{fig:real_datasets}
\end{figure}
\subsection{An Experiment}
To support our argument that the global-local low-rank structure would be a reasonable assumption,
we perform the following numerical experiment.
We take real HS images and numerically evaluate their global and low rank values.
The images come from six different datasets, namely, Chikusei, Cuprite, Indian Pines, University of Pavia, Washington DC Mall and Moffett Field.
They are shown in Fig.~\ref{fig:real_datasets} in color composite forms.
For each image, we obtain the local patches $\bX_i$'s by the equal-space rectangular segmentation in Fig.~\ref{fig:patches}.
For each local patch $\bX_i$ we evaluate its rank in an approximate manner, specifically, by finding the smallest integer $r$ such that
\[
\frac{ \sum_{j=1}^r \sigma_j(\bX_i)^2 }{ \sum_{j=1}^{\min\{M,L_i\}} \sigma_j(\bX_i)^2   } \geq 0.9999.
\]
Table~\ref{tab:local} shows the approximate ranks of the tested images under different patch sizes.
One can clearly see that all the tested images exhibit global-local low-rank characteristics.
For example, for the Chikusei image, the global rank is $10$ while the average local rank for $P= 16^2$ is around $6.5$.

\begin{table}[H]
\centering
\caption{The local approximate ranks of the real HS datasets. The patch size refers to the spectral-spatial dimension, $M \times L_i$, of the local patches.}
\label{tab:local}
\resizebox{.7\linewidth}{!}{%
\begin{tabular}{|c|c|c|c|c|c|}
\hline
\multirow{3}{*}{$P$} & \multicolumn{2}{c}{Chikusei}  & \multicolumn{2}{|c|}{Cuprite} \\
& \multicolumn{2}{c}{($480\times480$ pixels, $128$ bands)} & \multicolumn{2}{|c|}{($480\times480$ pixels, $187$ bands)}   \\
\cline{2-5}
 & patch size & approx. rank & patch size & approx. rank   \\\hline
$1^2$ & $128\times230,400$ & 10 & $187\times230,400$ & 9  \\
$2^2$ & $128\times57,600$ & 8.75$\pm$0.96 & $187\times57,600$ & 8.25$\pm$1.26  \\
$3^2$ & $128\times25,600$ & 8.33$\pm$1.73 & $187\times25,600$ & 7.44$\pm$1.33  \\
$4^2$ & $128\times14,400$ & 8.19$\pm$1.83 & $187\times14,400$ & 7.13$\pm$1.20  \\
$5^2$ & $128\times9,216$ & 7.92$\pm$1.89 & $187\times9,216$ & 6.76$\pm$1.13  \\
$6^2$ & $128\times6,400$ & 7.81$\pm$1.98 & $187\times6,400$ & 6.64$\pm$1.13  \\
$8^2$ & $128\times3,600$ & 7.38$\pm$2.13 & $187\times3,600$ & 6.38$\pm$1.06  \\
$10^2$ & $128\times2,304$ & 7.08$\pm$2.29 & $187\times2,304$ & 6.12$\pm$1.09   \\
$12^2$ & $128\times1,600$ & 6.89$\pm$2.34 & $187\times1,600$ & 6.03$\pm$1.02  \\
$15^2$ & $128\times1,024$ & 6.52$\pm$2.35 & $187\times1,024$ & 5.84$\pm$1.00   \\
$16^2$ & $128\times900$ & 6.47$\pm$2.36 & $187\times900$ & 5.79$\pm$1.01 \\
\hline
\multirow{3}{*}{$P$} & \multicolumn{2}{c}{Indian Pine} & \multicolumn{2}{|c|}{University of Pavia}  \\
& \multicolumn{2}{|c|}{($120\times120$ pixels, $178$ bands)} & \multicolumn{2}{|c|}{($240\times240$ pixels, $103$ bands)}  \\
\cline{2-5}
 & patch size & approx. rank & patch size & approx. rank \\\hline
$1^2$ & $178\times14,400$ & 22 & $103\times57,600$ & 22   \\
$2^2$ & $178\times3,600$ & 19.75$\pm$1.71 & $103\times14,400$ & 20.75$\pm$1.89  \\
$3^2$ & $178\times1,600$ & 17.56$\pm$4.10 & $103\times6,400$ & 20.44$\pm$2.07  \\
$4^2$ & $178\times900$ & 16.63$\pm$3.77 & $103\times3,600$ & 20.44$\pm$2.66  \\
$5^2$ & $178\times576$ & 15.20$\pm$3.94 & $103\times2,304$ & 20.28$\pm$2.84   \\
$6^2$ & $178\times400$ & 14.69$\pm$4.44 & $103\times1,600$ & 20.03$\pm$3.08  \\
$8^2$ & $178\times255$ & 13.36$\pm$4.47 & $103\times900$ & 19.80$\pm$3.43   \\\hline
\multirow{3}{*}{$P$} & \multicolumn{2}{c}{Washington DC Mall} & \multicolumn{2}{|c|}{Moffett Field}  \\
& \multicolumn{2}{|c|}{($240\times240$ pixels, $191$ bands)} &  \multicolumn{2}{|c|}{($240\times240$ pixels, $187$ bands)} \\
\cline{2-5}
 & patch size & approx. rank & patch size & approx. rank \\\hline
$1^2$ & $191\times57,600$ & 6 & $187\times57,600$ & 13 \\
$2^2$ & $191\times14,400$ & 5.50$\pm$0.58 & $187\times14,400$ & 12.00$\pm$2.71 \\
$3^2$ & $191\times6,400$ & 5.22$\pm$0.67 & $187\times6,400$ & 11.56$\pm$2.65 \\
$4^2$ & $191\times3,600$ & 4.50$\pm$1.03 & $187\times3,600$ & 11.00$\pm$2.94 \\
$5^2$ & $191\times2,304$ & 4.52$\pm$1.26 & $187\times2,304$ & 10.68$\pm$2.87 \\
$6^2$ & $191\times1,600$ & 4.39$\pm$1.23 & $187\times1,600$ & 10.56$\pm$2.92 \\
$8^2$ & $191\times900$ & 4.20$\pm$1.17 & $187\times900$ & 10.08$\pm$2.89 \\\hline
\end{tabular}
}
\end{table}

\subsection{The Global-Local Low-Rank Matrix Estimation Formulation}

Under the global-local low-rank assumption in the preceding subsection,
it is natural to formulate the HSR problem as the following global-local low-rank matrix estimation problem:
\begin{equation} \label{eq:prob_gloria_reg}
\min_{\bX \in \setX}  ~ \ell(\bX) + \sum_{i=0}^P \gamma_i \, {\rm rank}(\bX_i),
\end{equation}
where $\gamma_0,\gamma_1,\cdots,\gamma_P > 0$ are given regularization parameters;
we denote $\bX_0 = \bX$ for notational convenience;
$\ell$ has been defined in \eqref{eq:loss_fn};
$\setX \subseteq \Rbb^{M \times L}$ is given by $\setX = [ 0, 1 ]^{M \times L}$.

As reviewed previously, the standard approach to handle problem \eqref{eq:prob_rank_reg} is to approximate each ${\rm rank}(\bX_i)$ by the nuclear norm $\| \bX_i \|_*$.
Here we pursue a different option, namely, the the smooth Schatten-$p$ approximation \cite{mohan2012iterative}.
To put it into context, let us first define a notation.
Given a symmetric $n \times n$ positive definite matrix $\bA$ and a number $p$, we define
\begin{equation} \label{eq:mat_p}
\bA^p = \bU \bLam^p \bU^T,
\end{equation}
where $\bU$ and $\bLam$ constitute the eigen-decomposition $\bA  = \bU \bLam \bU^T$;
note that $\bU$ is orthogonal, $\bLam = \Diag(\lambda_1,\ldots,\lambda_n)$ with $\lambda_1 \geq \ldots \geq \lambda_n > 0$, and $\bLam^p = \Diag(\lambda_1^p,\ldots,\lambda_n^p)$.
The smooth Schatten-$p$ function of an $M \times L$ matrix $\bX$, with $M \leq L$, is defined as
\[
    \phi_{p,\tau}(\bX) = \sum_{i=1}^M \left( \sigma_i(\bX)^2 + \tau \right)^{p/2} = {\rm tr}\left(  (\bX \bX^T + \tau \bI )^{p/2}   \right),
\]
where $p > 0, \tau > 0$ are given.
This function has the following properties:
\begin{enumerate}[(i)]
\item $\phi_{p,\tau}$ is smooth (or has derivatives of all orders);
\item $\phi_{p,\tau}$ is convex for $p \geq 1$, and non-convex for $p < 1$;
\item as $\tau \rightarrow 0$, $\phi_{1,\tau}(\bX) \rightarrow \| \bX \|_*$;
\item as $p \rightarrow 0, \tau \rightarrow 0$, $\phi_{p,\tau}(\bX) \rightarrow {\rm rank}(\bX)$.
\end{enumerate}
As can be seen in the above properties, the smooth Schatten-$p$ function is a smooth approximation of rank.
Ideally we would like to choose very small $p$ and $\tau$ so that $\phi_{p,\tau}(\bX)$ closely approximates ${\rm rank}(\bX)$, but using very small $p$ and $\tau$ will also make $\phi_{p,\tau}$ poorly behaved (e.g., very large Lipschitz constant of the gradient of $\phi_{p,\tau}$).

By replacing ${\rm rank}(\bX_i)$ in problem \eqref{eq:prob_gloria_reg} with the smooth Schatten-$p$ function,
we obtain the Schatten-$p$ approximation of the global-local low-rank matrix estimation formulation \eqref{eq:prob_gloria_reg} as follows:
\begin{equation} \label{eq:prob_ssp_reg}
\min_{\bX \in \setX}  ~ \ell(\bX) + \sum_{i=0}^P \gamma_i \phi_{p,\tau}(\bX_i),
\end{equation}
where $p$ and $\tau$ are given.
We will be interested in the case of  $p < 1$.
The corresponding problem~\eqref{eq:prob_ssp_reg} is non-convex, but we found that, empirically, using $p < 1$ results in better recovery performance than using $p= 1$ (the smooth nuclear norm case).

\section{Global-Local Low-Rank Algorithm}
\label{sec:4}

In this section, we develop an algorithm for the global-local low-rank matrix estimation formulation \eqref{eq:prob_gloria_reg}.
Problem \eqref{eq:prob_gloria_reg} is a large-scale optimization problem with $ML$ optimization variables,
and computational efficiency is a main concern in algorithm design.
The algorithm to be presented is custom-designed for problem \eqref{eq:prob_gloria_reg}, where we exploit the problem structure for computational efficiency.
The main optimization concepts used in our algorithm design are majorization-minimization (MM) and the accelerated projected gradient method.

\subsection{Majorization-Minimization}

Firstly, we describe the MM method.
For notational convenience, let
\begin{equation} \label{eq:ef}
f(\bX) = \ell(\bX) + \sum_{i=0}^P \gamma_i \phi_{p,\tau}(\bX_i).
\end{equation}
In MM we seek a function $g(\bX;\bar{\bX})$, called a majorant of $f$, that satisfies
\begin{align*}
g(\bX;\bar{\bX}) \geq f(\bX),\quad
g(\bar{\bX};\bar{\bX}) = f(\bar{\bX}), \quad \forall \bX, \bar{\bX} \in \setX.
\end{align*}
We also require that, for any given $\bar{\bX} \in \setX$, $g(\cdot; \bar{\bX})$ is convex and continuously differentiable.
Given a starting point $\bX^0 \in \setX$,
the MM method handles problem \eqref{eq:prob_gloria_reg} by iteratively solving
\begin{equation} \label{eq:MM_iter}
\bX^{k+1} = \arg \min_{\bX \in \setX } \ g(\bX;\bX^{k}), \quad k=0,1,\ldots
\end{equation}
where the problem at each iteration in \eqref{eq:MM_iter} is a convex problem.
The MM iteration \eqref{eq:MM_iter} is known to guarantee convergence to a stationary solution to problem \eqref{eq:prob_gloria_reg} \cite{razaviyayn2013unified}.
We identify a majorant for problem \eqref{eq:prob_gloria_reg} by resorting to the following result.
\begin{Fact}{\cite[Sec.~3.1 and Appendix~A]{mohan2012iterative}} \label{fac:1}
	For $0 < p  \leq 1$, the smooth Schatten-$p$ function admits an alternative characterization
	\begin{equation} \label{eq:phi_alt}
	\phi_{p,\tau}(\bX) = \arg \min_{\bW \in \Sbb_{++}^M} \ \psi_{p,\tau}(\bX,\bW),
	\end{equation}
	where $\Sbb_{++}^M$ denotes the set of all $M \times M$ symmetric and positive definite matrices;
	\begin{equation*}\label{eq:sol}
        \psi_{p,\tau}(\bX,\bW) = \tfrac{p}{2} {\rm tr}(\bW(\bX \bX^T + \tau \bI)) + \tfrac{2-p}{p}{\rm tr}(\bW^{\frac{p}{p-2}}).
    \end{equation*}
	Also, the minimum in \eqref{eq:phi_alt} is uniquely attained at
	\[
	\bW^\star = (\bX \bX^T + \tau \bI)^{\frac{p}{2}-1}.
	\]
\end{Fact}
By applying Fact~\ref{fac:1} to \eqref{eq:ef},
and noting $\psi_{p,\tau}(\bX,\bW) \geq \psi_{p,\tau}(\bX,\bW^\star) =  \phi_{p,\tau}(\bX)$ for any $\bW \in \Sbb_{++}^M$,
we obtain the following majorant
\[
g(\bX;\bX^k) = \ell(\bX) + \sum_{i=0}^P \gamma_i \psi_{p,\tau}(\bX_i,\bW^k_i),
\]
where
\begin{equation} \label{eq:W_i_k}
\bW_i^k = (\bX_i^k (\bX_i^k)^T + \tau \bI)^{\frac{p}{2}-1}, \quad i=0,\ldots,P.
\end{equation}
Note that computing $\bW_i^k$ requires computing the eigendecomposition of $\bX_i^k (\bX_i^k)^T + \tau \bI$, which takes $\bigO(M^3)$ floating point operations; cf.~\eqref{eq:mat_p}.
It is easy to show that this $g$ is convex and continuously differentiable.
Also, solving the MM iteration $\min_{\bX \in \setX} g(\bX; \bX^k)$ in \eqref{eq:MM_iter} is the same as solving
\begin{equation} \label{eq:rls}
\min_{\bX \in \setX} \ \ell(\bX) + \sum_{i=0}^P \tfrac{p \gamma_i}{2} {\rm tr}(\bW_i^k \bX_i \bX_i^T ).
\end{equation}
The above problem is a quadratically regularized least-squares, with an iteratively reweighted quadratic regularizer.
Hence, the MM method developed above can be interpreted as an iteratively regularizer-reweighted least-squares method.
In fact, if we remove the bound constraints $\bX \in \setX$ we can solve problem~\eqref{eq:rls} in closed form.
However, we would like to keep the bound constraints, and this leads to the development in the next subsection.

\subsection{Accelerated Projected Gradient for Solving the MM Iteration}
\label{sec:APG}

Secondly, we describe an iterative method for solving the MM iteration in \eqref{eq:rls}.
We employ the accelerated projected gradient (APG) method \cite{nesterov1983method,beck2009fast,beck2017first}, which is a fast first-order algorithm.
The APG method for solving problem \eqref{eq:rls} is as follows.
Let
\begin{equation} \label{eq:gk}
g_k(\bX) = \ell(\bX) + \sum_{i=0}^P \tfrac{p \gamma_i}{2} {\rm tr}(\bW_i^k \bX_i \bX_i^T )
\end{equation}
for convenience.
Given a starting point $\bX^{0} \in \setX$, we perform the recursion
\begin{equation} \label{eq:apg_iter}
\bX^{j+1} = \Pi_\setX\left( \bZ^j - \tfrac{1}{\beta_j} \nabla g_k(\bZ^j) \right), \
j= 0,1,\ldots
\end{equation}
Here, $1/\beta_j > 0$ is the step size;
$\nabla g_k(\bX)$ denotes the gradient of $g_k$ at $\bX$;
$\Pi_\setX(\bY) = \arg \min_{\bX \in \setX} \| \bY- \bX \|_F^2$ denotes the projection  onto $\setX$;
$\bZ^j$ is the extrapolated point of the $\bX^j$'s and is given by
\begin{equation*} \label{eq:Z}
\bZ^j = \bX^j + \alpha_j (\bX^j - \bX^{j-1});
\end{equation*}
where $\bX^{-1} = \bX^0$;
the sequence $\{ \alpha_j \}$, called the extrapolation sequence, is given by
\begin{equation} \label{eq:extra_seq}
\alpha_j = \frac{\xi_{j-1}-1}{\xi_j},
\quad
\xi_j = \frac{1 + \sqrt{1+ 4\xi_{j-1}^2}}{2},
\end{equation}
with $\xi_{-1}= 0$.
The APG iteration \eqref{eq:apg_iter} is efficient to implement.
The gradient $\nabla g_k(\bX)$ is given by
\begin{align}
&\nabla g_k(\bX) \nonumber\\ = & \nabla \ell(\bX) +  \sum_{i=0}^P \tfrac{p\gamma_i}{2} \nabla \left( {\rm tr}(\bW_i^k \bX_i \bX_i^T ) \right) \nonumber  \\
 = & \nabla \ell(\bX) + p ( \gamma_0 \bW^k_0 \bX + [~ \gamma_1 \bW_1^k \bX_1, \ldots, \gamma_P \bW_P^k \bX_P  ~]).
\label{eq:nabla_gk}
\end{align}
Also, we have
\begin{align} \label{eq:nabla_el}
\nabla \ell(\bX)  & = \bF^T \bF \bX - \bF^T \bYm + \bX \bG \bG^T - \bYh \bG^T.
\end{align}
It can be verified that, given $\bW_0^k, \ldots, \bW_P^k$, computing \eqref{eq:nabla_gk}--\eqref{eq:nabla_el} takes $\bigO(M(ML+ {\rm nnz}(\bG)))$ floating-point operations,
where ${\rm nnz}(\bG)$ denotes the number of nonzero elements of $\bG$.
We should note that a large number of elements of $\bG$ are zero.
Recall that $\bG$ describes the spatial degradation process of local spatial blurring and downsampling.
One can show that the number of nonzero elements of each column of $\bG$ depends on the size of the blurring kernel,
and in practice the blurring kernel size is often small.
Also, the projection operation $\Pi_\setX$ for the bound set $\setX= [0,1]^{M \times L}$ is merely a clipping function, i.e.,
\begin{equation} \label{eq:Pi}
\Pi_\setX(\bY) = \max\{ \bzero, \min\{ \bone, \bY \} \},
\end{equation}
where $\bzero$ and $\bone$ denote all-zero and all-one matrices, resp., and $\min$ and $\max$ are taken in the element-wise manner.

We complete the APG development by specifying our step-size rule.
The APG method guarantees convergence to the optimal solution if $\beta_k$ is chosen to be a Lipschitz constant of $\nabla g_k$ \cite{beck2009fast,beck2017first}. We have the following result.
\begin{Fact} \label{fac:lip}
	A Lipschitz constant of $\nabla g_k$ in \eqref{eq:gk} is
	\begin{align} \label{eq:lip}
	L_{g_k} = & \lammax(\bF^T \bF+ p \gamma_0 \bW^k_0) + \lammax( \bG\bG^T) \nonumber\\
    &\qquad\qquad\qquad\qquad + p \max_{i=1,\ldots,P} \gamma_i \lammax(\bW_i^k),
	\end{align}
	where $\lammax(\cdot)$ denotes the largest eigenvalue of the argument.
\end{Fact}
The proof of Fact~\ref{fac:lip} is shown in Appendix~A.
The computational cost of \eqref{eq:lip} is mainly with computing the largest eigenvalues of $M \times M$ symmetric matrices,
and there are $P+2$ such eigenvalues to compute.
The eigenvalue $\lammax( \bG\bG^T)$ can be computed offline, and the eigenvalues $\lammax(\bW_i^k)$ are byproducts of computing $\bW_i^k$ in~\eqref{eq:W_i_k}
(again, cf. \eqref{eq:mat_p}).
It suffices to calculate $\lambda_{\max}(\bF^T\bF+p\gamma_0\bW_0^k)$ at every MM iteration.
Hence, computing \eqref{eq:lip} takes $\bigO(M^3)$ floating-point operations.
\vspace{-.5em}
\subsection{Algorithm Speedup via Inexact MM}

Let us summarize the MM method developed in the last two subsections:
We perform the MM iteration \eqref{eq:MM_iter}.
Every iteration requires solving a regularized least-squares (with bound constraints) exactly, and we do that by applying the APG method in \eqref{eq:apg_iter}--\eqref{eq:Pi}.
The setback with this exact MM approach is that while the APG method is considered fast, it still takes time to solve each MM problem exactly.

The algorithm we finally adopt is an inexact MM method.
At each MM iteration, we apply the APG method with one iteration only.
Specifically, we replace the exact MM iteration \eqref{eq:MM_iter} by
\begin{subequations} \label{eq:iMM_iter}
	\begin{align}
	\bX^{k+1} & = \Pi_\setX\left( \bZ^k - \tfrac{1}{\beta_k} \nabla g_k(\bZ^k) \right), \quad k=0,1,\ldots \\
	\bZ^k & = \bX^k + \alpha_k (\bX^k - \bX^{k-1}),
	\end{align}
\end{subequations}
where $\alpha_k$ is given by \eqref{eq:extra_seq};
$\beta_k$ is chosen as $\beta_k = 1/L_{g_k}$, with $L_{g_k}$ given by \eqref{eq:lip}.
By using this inexact MM update we hope that the total number of iterations (i.e., the sum of APG iterations incurred by all the MM iterations) may be reduced, and the runtime improved.
Based on our numerical experience, the inexact MM method runs much faster than the exact MM.
Also,
it is shown in \cite{shao2018framework} that, under some technical assumptions, the above inexact MM method guarantees convergence to a stationary solution.

We summarize the inexact MM algorithm in pseudo-form in Algorithm~\ref{alg:GLORIA}. We call this algorithm  Global-Local lOw-Rank promotIng Algorithm (GLORIA).
It can be verified that the complexity of GLORIA is $\bigO(M(ML+PM^2 + {\rm nnz}(\bG)))$ per iteration.
\begin{algorithm}[H]
\caption{GLORIA}
\label{alg:GLORIA}
{\footnotesize
    \begin{algorithmic}[1]
        \State {\bf input} a starting point $\bX^0$
        \State $\bX^{-1}=\bX^0,\xi_0=0$;
        \State $k = 0$;
        \Repeat
        \State $\xi_{k+1}=\left(1+\sqrt{1+4\xi_k}\right)/2$;
        \State $\alpha_k=(\xi_k-1)/\xi_{k+1}$;
        \State $\bZ^k = \bX^k+\alpha_k(\bX^k-\bX^{k-1})$;
        \State $\nabla\ell(\bZ^k)=\bF^T\bF\bZ^k-\bF^T\bYm+\bZ^k\bG\bG^T-\bYh\bG^T$;
        \State
        compute the eigendecomposition $ \bU_i \bLam_i \bU_i^T:= \bZ^k_i(\bZ^k_i)^T+\tau\bI$,~ $i=0,1,\ldots,P$;
        \State
        $\bW^k_i = \bU_i  \bLam_i^{\frac{p}{2}-1} \bU_i^T$,~ $i=0,1,\ldots,P$;
        \State
        $\nabla g_k(\bZ^k)=\newline{\color{white}.}~~\qquad\qquad\nabla \ell(\bZ^k)+ p \gamma_0 \bW_0^k \bZ_0^k + p [~ \gamma_1 \bW_1^k \bZ^k_1,\ldots,\gamma_P \bW_P^k \bZ^k_P ~]$;
        \State
        $L_{g_k} = \newline{\color{white}.}\lambda_{\max}(\bF^T\bF+ p \gamma_0  \bW_0^k )+\lambda_{\max}(\bG\bG^T) + p \max_{i=1,\ldots,P} \gamma_i
        \lambda_{M,i}^{\frac{p}{2}-1}$;
        \State $\tilde{\bX}^{k+1} = \bZ^k-(1/L_{g_k})\nabla g_k(\bZ^k)$;
        \State $\bX^{k+1} = \max\{ \bzero, \min\{ \bone, \tilde{\bX}^{k+1} \} \}$;
        \State $k:=k+1$;
        \Until{some stopping criterion is satisfied.}
        \State {\bf output} $\bX_{{\rm est}}=\bX^k$.
    \end{algorithmic}}
\end{algorithm}

\section{Numerical Results}
\label{sec:experiment}
We performed extensive numerical experiments to benchmark GLORIA against a number of existing algorithms.
The benchmarked algorithms we choose are considered most representative in the context or are related to our method.
Namely, they are
GSA~\cite{aiazzi2007improving}, GLP~\cite{aiazzi2006mtf}, CNMF~\cite{yokoya2012coupled}, FUMI~\cite{wei2016multiband}, HySure~\cite{simoes2015convex},
the nuclear norm minimization (NNM) method in \eqref{eq:prob_mn_reg},
and LRSR~\cite{dian2018hyperspectral}.
GSA and GLP are pansharpening-based methods;
CNMF and FUMI are representative methods in matrix factorization-based HSR;
HySure is a representative method in dictionary-based regression, and it employs spatial total variation regularization in its regression;
LRSR is another dictionary-based regression method which uses local nuclear-norm regularization;
NNM is regarded as the baseline method for low-rank matrix estimation in the context of matrix completion, and thus we include it in our experiments.
All the algorithms are implemented on a desktop computer with Intel Core i7-5760X@3GHz CPU and 32GB memory. Codes are written in MATLAB R2015a.

The parameter settings of GLORIA are as follows, unless specified otherwise.
The parameters of the smooth Schatten-$p$ function are $p= 1/2$, $\tau= 1$.
The local patches are obtained by dividing the image into equal-spaced rectangular blocks, as in Fig.~\ref{fig:patches}.
The regularization parameters $\gamma_i$'s are chosen to be identical; i.e., $\gamma_0 = \gamma_1 = \cdots = \gamma_P = \gamma$.
We initialize the algorithm by using a $[0,1]$-uniform i.i.d. generated $\bX$.
We stop the algorithm when the relative change of the objective function is below $10^{-5}$ or when the number of iterations exceeds $100$.

The parameter settings of the benchmarked algorithms basically follow the recommended settings in~\cite{yokoya2017hyperspectral,dian2018hyperspectral}.
In addition, for matrix factorization and dictionary-based regression methods, we fix the target rank as $N= 30$.
We use VCA to initialize the matrix factorization algorithms, and we stop the algorithms by the same stopping rule as that for GLORIA.
Also, the NNM method is implemented by applying the APG method in \cite{ryan2018b} to problem \eqref{eq:prob_mn_reg}.

The performance measures employed for evaluating the recovery performance are the peak SNR (PSNR), spectral angle mapper (SAM), Erreur Relative Globale Adimensionnelle de Sythe\`{s}e (ERGAS) and universal image quality index (UIQI).
They have been extensively used in the HSR literature, and we refer the reader to \cite{loncan2015hyperspectral} for their definitions.

\subsection{Semi-Real Data Experiments}

First, we consider semi-real data experiments.
The experiments were based on the widely-used Wald's protocol~\cite{wald2000quality}, where we take a real HS image as the ground-truth SR image $\bX$ and use it to generate the observed MS and HS images, $\bYh$ and $\bYm$, through the  model~\eqref{eq:sig_mod}.
We consider the following real HS datasets.
\begin{enumerate}[1.]
    \item
    {\it Hyperspec-VNIR-C Chikusei:}
    This dataset was acquired by the Headwall Hyperspec-VNIR-C imaging sensor \cite{NYokoya2016}. It covers 128 spectral bands whose wavelength range is from 363nm to 1,018nm.
    We take a 480$\times$480 subimage from this dataset as our SR image.
    \item
    {\it AVIRIS Indian Pines:}
    This dataset was captured by the Airborne Visible/Infrared Imaging Spectrometer (AVIRIS) HS sensor \cite{vane1993airborne}.
    The wavelength range is from 400nm to 2,500nm.
    It has 178 spectral bands after dropping bands that are corrupted by water absorption.
    In the experiment, a 120$\times$120 subimage is used.
    \item
    {\it HYDICE Washington DC Mall:}
    This dataset was taken by the Hyperspectral Digital Imagery Collection Experiment (HYDICE) HS sensor~\cite{basedow1995hydice}.
    We take a subimage of this dataset, which has 240$\times$240 pixels and 191 clean spectral bands. The wavelength range is from 400nm to 2,500nm.
    \item
    {\it ROSIS University of Pavia:}
    This dataset was measured by the Reflective Optics System Imaging Spectrometer (ROSIS) HS sensor.
    This dataset has 103 spectral bands whose wavelength range is from 430nm to 850nm.
    We take a 240$\times$240 subimage from this dataset as the SR image.
    \item
    {\it AVIRIS Moffett Field:}
    This dataset, recorded by AVIRIS HS sensor, has 187 uncorrupted spectral bands.
    The wavelength range is from 400nm to 2,500nm.
    We take a 240$\times$240 subimage from this dataset.
\end{enumerate}
These five images have been displayed in Fig.~\ref{fig:real_datasets}.
The settings with the spectral and spatial measurement response matrices $\bF$ and $\bG$ should also be described.
The matrix $\bF$ is chosen such that it is equivalent to the spectral response of either the Landsat 4 TM sensor~\cite{chander2009summary} (6 bands, with spectral coverage from 400nm to 2,500nm) or the IKONOS sensor~\cite{dial2003ikonos} (4 bands, with spectral coverage from 450nm to 900nm).
We choose the Landsat 4 TM sensor response for Indian Pines, Washington DC Mall and Moffett Field, and the IKONOS sensor for Chikusei and University of Pavia;
such choosing is to match the spectral coverage of the HS images.
As discussed previously, $\bG$ corresponds to the process of spatial blurring and downsampling.
The blurring function is a 11$\times$11 Gaussian point spread function, with variance $1.7^2$.
The downsampling is done every $4$ pixels, both horizontally and vertically.
Furthermore, the noise terms $\bVh$ and $\bVm$ are randomly generated following an i.i.d. mean-zero Gaussian distribution.
We fix the SNRs at ${\sf SNR}_{\rm M}={\sf SNR}_{\rm H}=25{\rm dB}$.

\begin{figure}[!bp]
\begin{minipage}[c]{.98\textwidth}
    \begin{minipage}[c]{.243\textwidth}
        \centering
        \includegraphics[width=\linewidth]{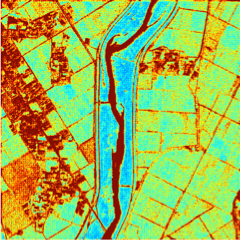}
        (a) GSA
    \end{minipage}
    \begin{minipage}[c]{.243\textwidth}
        \centering
        \includegraphics[width=\linewidth]{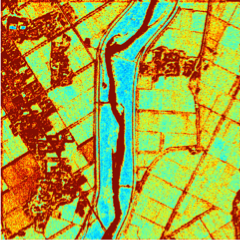}
        (b) GLP
    \end{minipage}
    \begin{minipage}[c]{.243\textwidth}
        \centering
        \includegraphics[width=\linewidth]{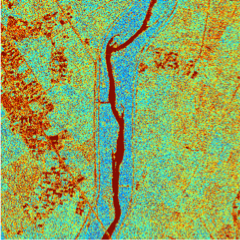}
        (c) CNMF
    \end{minipage}
    \begin{minipage}[c]{.243\textwidth}
        \centering
        \includegraphics[width=\linewidth]{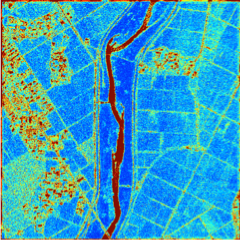}
        (d) FUMI
    \end{minipage}
    \begin{minipage}[c]{.243\textwidth}
        \centering
        \includegraphics[width=\linewidth]{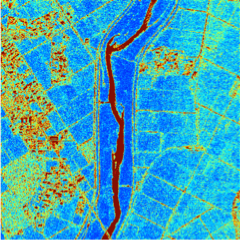}
        (e) HySure
    \end{minipage}
    \begin{minipage}[c]{.243\textwidth}
        \centering
        \includegraphics[width=\linewidth]{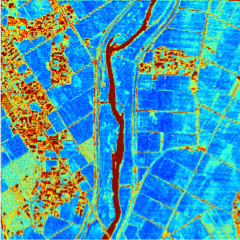}
        (f) LRSR
    \end{minipage}
    \begin{minipage}[c]{.243\textwidth}
        \centering
        \includegraphics[width=\linewidth]{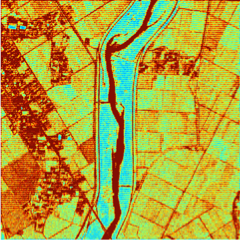}
        (g) NNM
    \end{minipage}
    \centering
    \begin{minipage}[c]{.243\textwidth}
        \centering
        \includegraphics[width=\linewidth]{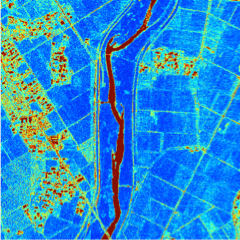}
        (h) GLORIA
    \end{minipage}
\end{minipage}
\begin{minipage}[c]{.01\textwidth}
    \centering
    \includegraphics[height=51\linewidth]{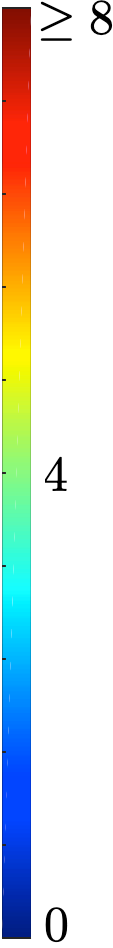}
    \\\vspace*{1.3em}
\end{minipage}
    \caption{SAM maps of the Chikusei dataset.}
    \label{fig:sam_Chikusei}\vspace*{1em}
\centering
\begin{minipage}[c]{.98\textwidth}
    \begin{minipage}[c]{.243\textwidth}
        \centering
        \includegraphics[width=\linewidth]{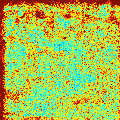}
        (a) GSA
    \end{minipage}
    \begin{minipage}[c]{.243\textwidth}
        \centering
        \includegraphics[width=\linewidth]{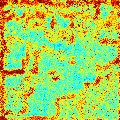}
        (b) GLP
    \end{minipage}
    \begin{minipage}[c]{.243\textwidth}
        \centering
        \includegraphics[width=\linewidth]{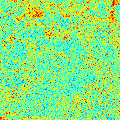}
        (c) CNMF
    \end{minipage}
    \begin{minipage}[c]{.243\textwidth}
        \centering
        \includegraphics[width=\linewidth]{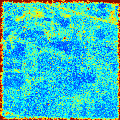}
        (d) FUMI
    \end{minipage}
    \begin{minipage}[c]{.243\textwidth}
        \centering
        \includegraphics[width=\linewidth]{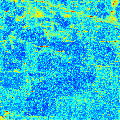}
        (e) HySure
    \end{minipage}
    \begin{minipage}[c]{.243\textwidth}
        \centering
        \includegraphics[width=\linewidth]{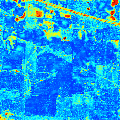}
        (f) LRSR
    \end{minipage}
    \begin{minipage}[c]{.243\textwidth}
        \centering
        \includegraphics[width=\linewidth]{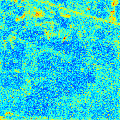}
        (g) NNM
    \end{minipage}
    \centering
    \begin{minipage}[c]{.243\textwidth}
        \centering
        \includegraphics[width=\linewidth]{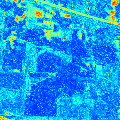}
        (h) GLORIA
    \end{minipage}
\end{minipage}
\begin{minipage}[c]{.01\textwidth}
    \centering
    \includegraphics[height=51\linewidth]{figs/bar.png}
    \\\vspace*{1.3em}
\end{minipage}
    \caption{SAM maps of the Indian Pines dataset.}
    \label{fig:sam_Indianpines}
\end{figure}

\begin{figure}[bp!]
\centering
\begin{minipage}[c]{.98\textwidth}
   \begin{minipage}[c]{.243\textwidth}
        \centering
        \includegraphics[width=\linewidth]{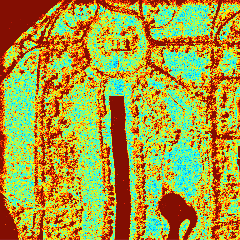}
        (a) GSA
    \end{minipage}
    \begin{minipage}[c]{.243\textwidth}
        \centering
        \includegraphics[width=\linewidth]{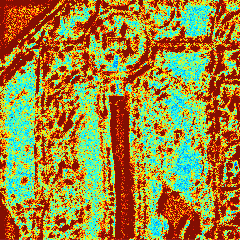}
        (b) GLP
    \end{minipage}
    \begin{minipage}[c]{.243\textwidth}
        \centering
        \includegraphics[width=\linewidth]{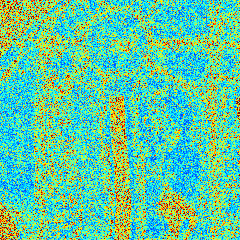}
        (c) CNMF
    \end{minipage}
    \begin{minipage}[c]{.243\textwidth}
        \centering
        \includegraphics[width=\linewidth]{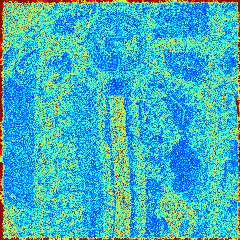}
        (d) FUMI
    \end{minipage}
    \begin{minipage}[c]{.243\textwidth}
        \centering
        \includegraphics[width=\linewidth]{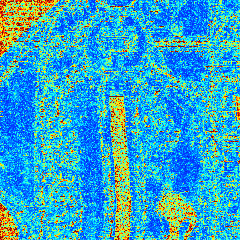}
        (e) HySure
    \end{minipage}
    \begin{minipage}[c]{.243\textwidth}
        \centering
        \includegraphics[width=\linewidth]{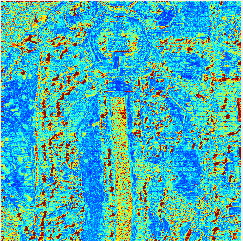}
        (f) LRSR
    \end{minipage}
    \begin{minipage}[c]{.243\textwidth}
        \centering
        \includegraphics[width=\linewidth]{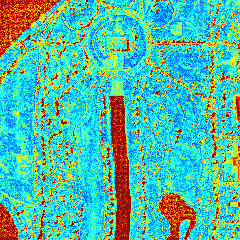}
        (g) NNM
    \end{minipage}
    \centering
    \begin{minipage}[c]{.243\textwidth}
        \centering
        \includegraphics[width=\linewidth]{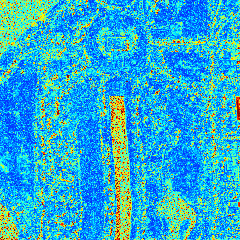}
        (h) GLORIA
    \end{minipage}
\end{minipage}
\begin{minipage}[c]{.01\textwidth}
    \centering
    \includegraphics[height=51\linewidth]{figs/bar.png}
    \\\vspace*{1.3em}
\end{minipage}
    \caption{SAM maps of the algorithms, the Washington DC Mall dataset.}
    \label{fig:sam_WashingtonDC}\vspace*{1em}
\begin{minipage}[c]{.98\textwidth}
    \begin{minipage}[c]{.243\textwidth}
        \centering
        \includegraphics[width=\linewidth]{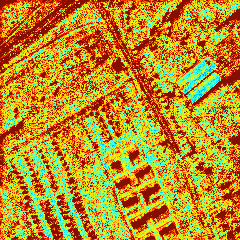}
        (a) GSA
    \end{minipage}
    \begin{minipage}[c]{.243\textwidth}
        \centering
        \includegraphics[width=\linewidth]{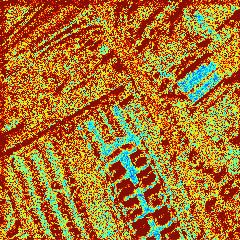}
        (b) GLP
    \end{minipage}
    \begin{minipage}[c]{.243\textwidth}
        \centering
        \includegraphics[width=\linewidth]{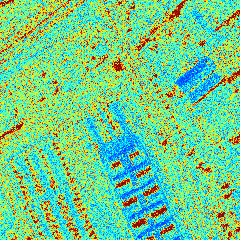}
        (c) CNMF
    \end{minipage}
    \begin{minipage}[c]{.243\textwidth}
        \centering
        \includegraphics[width=\linewidth]{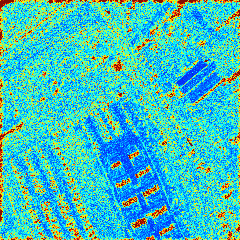}
        (d) FUMI
    \end{minipage}
    \begin{minipage}[c]{.243\textwidth}
        \centering
        \includegraphics[width=\linewidth]{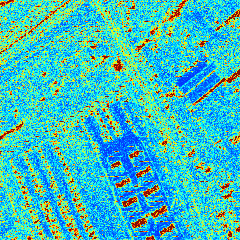}
        (e) HySure
    \end{minipage}
    \begin{minipage}[c]{.243\textwidth}
        \centering
        \includegraphics[width=\linewidth]{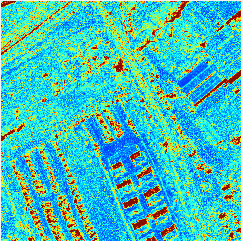}
        (f) LRSR
    \end{minipage}
    \begin{minipage}[c]{.243\textwidth}
        \centering
        \includegraphics[width=\linewidth]{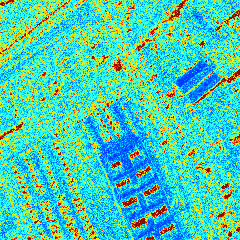}
        (g) NNM
    \end{minipage}
    \centering
    \begin{minipage}[c]{.243\textwidth}
        \centering
        \includegraphics[width=\linewidth]{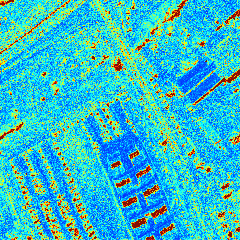}
        (h) GLORIA
    \end{minipage}
\end{minipage}
\begin{minipage}[c]{.01\textwidth}
    \centering
    \includegraphics[height=51\linewidth]{figs/bar.png}
    \\\vspace*{1.3em}
\end{minipage}
    \caption{SAM maps of the  University of Pavia dataset.}
    \label{fig:sam_UnivPavia}
\end{figure}

\begin{figure}[H]
\centering
\begin{minipage}[c]{.98\textwidth}
    \begin{minipage}[c]{.243\textwidth}
        \centering
        \includegraphics[width=\linewidth]{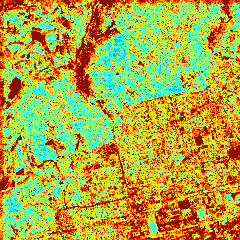}
        (a) GSA
    \end{minipage}
    \begin{minipage}[c]{.243\textwidth}
        \centering
        \includegraphics[width=\linewidth]{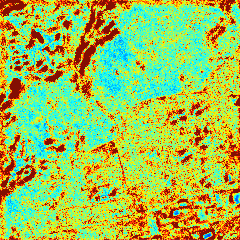}
        (b) GLP
    \end{minipage}
    \begin{minipage}[c]{.243\textwidth}
        \centering
        \includegraphics[width=\linewidth]{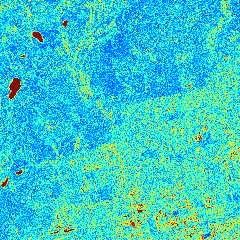}
        (c) CNMF
    \end{minipage}
    \begin{minipage}[c]{.243\textwidth}
        \centering
        \includegraphics[width=\linewidth]{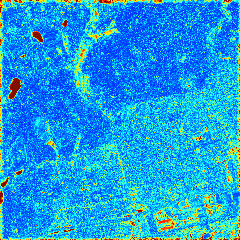}
        (d) FUMI
    \end{minipage}
    \begin{minipage}[c]{.243\textwidth}
        \centering
        \includegraphics[width=\linewidth]{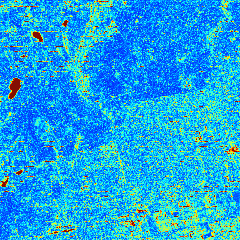}
        (e) HySure
    \end{minipage}
    \begin{minipage}[c]{.243\textwidth}
        \centering
        \includegraphics[width=\linewidth]{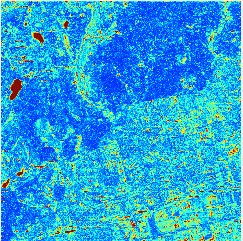}
        (f) LRSR
    \end{minipage}
    \begin{minipage}[c]{.243\textwidth}
        \centering
        \includegraphics[width=\linewidth]{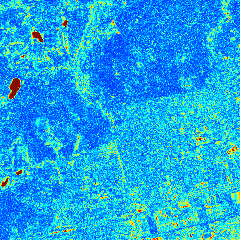}
        (g) NNM
    \end{minipage}
    \centering
    \begin{minipage}[c]{.243\textwidth}
        \centering
        \includegraphics[width=\linewidth]{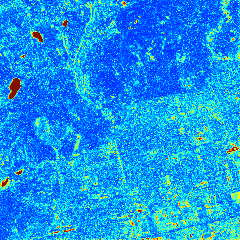}
        (h) GLORIA
    \end{minipage}
\end{minipage}
\begin{minipage}[c]{.01\textwidth}
    \centering
    \includegraphics[height=51\linewidth]{figs/bar.png}
    \\\vspace*{1.3em}
\end{minipage}
    \caption{SAM maps of the Moffett Field dataset.}
    \label{fig:sam_Moffett}
\end{figure}

Table~\ref{tab:implementation} summarizes some of the settings with the experiments.
There, we also show the settings with the patch number $P$ of GLORIA.
The regularization parameter of GLORIA is fixed as $\gamma=20/({\sf SNR}_{\rm M}  + {\sf SNR}_{\rm H})$.
We ran $50$ independent trials in each image.
The obtained performance is shown in  Table~\ref{table:semi_real_1}, where, for each performance measure, we use blue, brown and red boldfaced letters to mark the best, second best and third best algorithms.
To give the reader an additional reference on the performance, we show the SAM maps of the various algorithms in Figs.~\ref{fig:sam_Chikusei}--\ref{fig:sam_Moffett};
note that the SAM maps shown are from one realization.

From Table~\ref{table:semi_real_1} we see that, except for runtimes, GLORIA ranks best or second best in all of the performance measures.
From Figs.~\ref{fig:sam_Chikusei}--\ref{fig:sam_Moffett}, we also see that GLORIA yields good results compared to the other algorithms---and
this is particularly so for the Indian Pine image in Fig.~\ref{fig:sam_Indianpines}.
In fact, we observe that even the NNM method, which is the baseline low-rank matrix estimation method and can be regarded as the precursor of our global-local low-rank pursuit,  works reasonably.
The above reported results suggest that the exploitation of low-rank spectral-spatial data structures in HSR is a working idea.
We speculate that the good performance of GLORIA compared to the other algorithms is because GLORIA exploits the local low-rank data structure, which may provide better robustness to the EV effects.
We will use synthetic experiments to examine the EV effects in the next subsection.

We should also discuss the runtime performance.
The best algorithms are GSA and GLP, which do not perform very well in recovery performance.
Let us compare GLORIA to the representative CNMF and HySure methods.
GLORIA runs faster than CNMF and HySure for the cases of Chikusei and University of Pavia, and slower for the cases of Washington DC Mall and Moffett Field.
On this issue, we should note that GLORIA deals with $ML$ optimization variables, e.g., $29,491,200$ variables in Chikusei.
In comparison, CNMF and HySure require $(M+L)N$ and $NL$ optimization variables, resp., which amount to $6,915,840$ and $6,912,000$ variables, resp., in Chikusei.
In terms of runtime per variable, GLORIA is considered efficient.

\begin{table}[H]
\centering
\caption{Some settings with the semi-real data experiments.}
\label{tab:implementation}
\resizebox{\textwidth}{!}{%
\begin{tabular}{|c|c|c|c|c|c|}\hline
& Chikusei & Indian Pines & Washinton DC Mall & University of Pavia & Moffett Field  \\\hline
MS response & IKONOS & Landsat 4 TM & Landsat 4 TM & IKONOS & Landsat 4 TM \\\hline
Image size ($M\times L$)  & 128$\times$230,400 & 178$\times$14,400 & 191$\times$57,600 & 103$\times$57,600 & 187$\times$57,600 \\\hline
Patch size ($M\times L_i$) & 128$\times$3,600 & 178$\times$900 & 190$\times$3,600 & 103$\times$3,600 & 187$\times$3,600 \\\hline
Patch number $P$ & 64 & 16 & 16 & 16 & 16 \\
\hline
\end{tabular}%
}
\end{table}

\begin{table}[!]
\centering
\caption{Average performance of the algorithms on semi-real datasets.}
\label{table:semi_real_1}
\resizebox{.83\linewidth}{!}{
\begin{tabular}{|c|c|c|c|c|c|}
\hline
Method & Time (sec.) & PSNR & SAM & ERGAS & UIQI \\\cline{1-6}
Ideal value & 0 & $\infty$ & 0 & 0 & 1 \\\hline
\multicolumn{6}{|c|}{Dataset - Chikusei}\\\hline
GSA & {\bf\blue1.44$\pm$0.15} & 33.11$\pm$0.01 & 4.85$\pm$0.01 & 6.59$\pm$0.01 & 0.787$\pm$0.000 \\
GLP & {\bf\brown43.44$\pm$8.84} & 29.13$\pm$0.00 & 5.36$\pm$0.01 & 7.92$\pm$0.01 & 0.733$\pm$0.000 \\
CNMF & 147.54$\pm$64.93 & 34.13$\pm$0.11 & 4.21$\pm$0.08 & 5.24$\pm$0.11 & 0.811$\pm$0.004 \\
FUMI & 293.31$\pm$24.18 & 35.18$\pm$0.04 & {\bf\brown2.55$\pm$0.03} & {\bf\brown4.03$\pm$0.03} & {\bf\brown0.882$\pm$0.001} \\
HySure & 130.88$\pm$11.80 & {\brown\bf36.15$\pm$0.15} & {\red\bf2.84$\pm$0.09} & {\red\bf4.31$\pm$0.08} & 0.863$\pm$0.003 \\
LRSR & 279.26$\pm$43.04 & {\bf\red35.84$\pm$0.09} & 2.88$\pm$0.03 & 4.41$\pm$0.06 & {\bf\red0.880$\pm$0.002} \\
NNM & {\bf\red91.95$\pm$12.48} & 32.63$\pm$0.01 & 4.96$\pm$0.00 & 7.49$\pm$0.01 & 0.764$\pm$0.000 \\
GLORIA & 119.99$\pm$21.78 & {\blue\bf37.73$\pm$0.01} & {\blue\bf2.31$\pm$0.00} & {\blue\bf3.59$\pm$0.00} & {\bf\blue0.894$\pm$0.000}  \\\hline\hline
\multicolumn{6}{|c|}{Dataset - Indian Pines}\\\hline
GSA & {\blue\bf0.44$\pm$0.08} & 19.07$\pm$2.24 & 7.19$\pm$9.30 & 4.09$\pm$2.78 & 0.512$\pm$0.067 \\
GLP & {\bf\brown2.38$\pm$0.51} & 18.06$\pm$0.48 & 4.84$\pm$0.09 & 3.66$\pm$0.20 & 0.397$\pm$0.034 \\
CNMF & 16.57$\pm$4.04 & 22.64$\pm$0.22 & 4.29$\pm$0.10 & 2.43$\pm$0.07 & 0.523$\pm$0.009 \\
FUMI & 11.40$\pm$0.30 & 24.85$\pm$0.15 & {\bf\red2.69$\pm$0.05} & 1.76$\pm$0.04 & {\bf\red0.762$\pm$0.007} \\
HySure & 10.96$\pm$0.37 & {\red\bf26.70$\pm$0.16} & 2.74$\pm$0.04 & {\bf\red1.41$\pm$0.03} & 0.698$\pm$0.008 \\
LRSR & 18.80$\pm$0.28 & {\bf\brown27.98$\pm$0.14} & {\bf\brown2.53$\pm$0.03} & {\bf\brown1.18$\pm$0.02} & {\bf\brown0.796$\pm$0.004} \\
NNM & {\red\bf6.81$\pm$0.89} & 26.43$\pm$0.04 & 2.93$\pm$0.01 & 1.71$\pm$0.01 & 0.679$\pm$0.002 \\
GLORIA & 14.56$\pm$0.92 & {\blue\bf29.09$\pm$0.03} & {\blue\bf2.28$\pm$0.00} & {\blue\bf1.04$\pm$0.00} & {\blue\bf0.804$\pm$0.001} \\
\hline\hline
\multicolumn{6}{|c|}{Dataset - Washington DC Mall}\\\hline
GSA & {\blue\bf0.97$\pm$0.09} & 20.09$\pm$0.10 & 6.51$\pm$0.03 & 20.35$\pm$0.38 & 0.654$\pm$0.004 \\
GLP & {\bf\brown7.68$\pm$1.83} & 18.20$\pm$0.09 & 7.00$\pm$0.09 & 14.85$\pm$0.06 & 0.582$\pm$0.005 \\
CNMF & 31.74$\pm$16.95 & 25.71$\pm$0.19 & 3.84$\pm$0.08 & 6.94$\pm$0.22 & 0.772$\pm$0.007 \\
FUMI & 44.33$\pm$0.69 & {\bf\brown29.17$\pm$0.10} & {\blue\bf2.49$\pm$0.02} & {\blue\bf2.91$\pm$0.06} & {\blue\bf0.915$\pm$0.004} \\
HySure & 31.28$\pm$8.81 & {\bf\red28.82$\pm$0.33} & {\bf\red3.00$\pm$0.07} & {3.74$\pm$0.25} & {\bf\red0.872$\pm$0.009} \\
LRSR & 71.92$\pm$1.55 & 27.91$\pm$0.16 & 3.85$\pm$0.10 & {\bf\red3.59$\pm$0.15} & 0.869$\pm$0.007 \\
NNM & {\bf\red27.63$\pm$1.71} & 27.57$\pm$0.12 & 3.28$\pm$0.02 & 8.05$\pm$0.03 & 0.813$\pm$0.005 \\
GLORIA & 57.06$\pm$8.14 & {\blue\bf29.36$\pm$0.23} & {\bf\brown2.85$\pm$0.02} & {\bf\brown3.18$\pm$0.20} & {\bf\brown0.888$\pm$0.002} \\
\hline\hline
\multicolumn{6}{|c|}{Dataset - University of Pavia}\\\hline
GSA & {\bf\blue 0.49$\pm$0.05} & 30.50$\pm$0.02 & 6.79$\pm$0.03 & 3.05$\pm$0.01 & 0.900$\pm$0.001 \\
GLP & {\bf\brown4.45$\pm$1.14} & 25.71$\pm$0.07 & 7.88$\pm$0.11 & 5.54$\pm$0.06 & 0.764$\pm$0.005 \\
CNMF & 27.08$\pm$21.05 & 35.86$\pm$0.18 & 4.09$\pm$0.09 & 1.84$\pm$0.04 & 0.946$\pm$0.002 \\
FUMI & 36.72$\pm$0.81 & 35.29$\pm$0.04 & {\blue\bf3.12$\pm$0.01} & 1.76$\pm$0.01 & {\bf\brown0.962$\pm$0.000} \\
HySure & 27.46$\pm$9.73 & {\bf\red36.71$\pm$0.14} & {\bf\red3.34$\pm$0.02} & {\red\bf1.55$\pm$0.01} & {\bf\red0.961$\pm$0.001} \\
LRSR & 58.29$\pm$1.20 & 36.43$\pm$0.09 &3.45$\pm$0.05 & 1.65$\pm$0.03 & 0.959$\pm$0.001 \\
NNM & {\bf\red10.55$\pm$0.41} & {\brown\bf36.81$\pm$0.01} & 3.48$\pm$0.00 & {\bf\brown1.55$\pm$0.00} & 0.957$\pm$0.000 \\
GLORIA & 16.68$\pm$0.50 & {\blue\bf37.51$\pm$0.01} & {\bf\brown3.16$\pm$0.00} & {\blue\bf1.46$\pm$0.00} & {\blue\bf0.964$\pm$0.000}
\\\hline\hline
\multicolumn{6}{|c|}{Dataset - Moffett Field}\\\hline
GSA & {\blue\bf0.87$\pm$0.06} & 30.56$\pm$0.02 & 5.84$\pm$0.01 & 2.79$\pm$0.01 & 0.843$\pm$0.000 \\
GLP & {\bf\brown6.50$\pm$0.70} & 28.18$\pm$0.19 & 5.30$\pm$0.08 & 3.53$\pm$0.07 & 0.775$\pm$0.013 \\
CNMF & {\bf\red17.38$\pm$9.60} & 36.08$\pm$0.12 & 3.19$\pm$0.06 & 1.57$\pm$0.03 & 0.919$\pm$0.002 \\
FUMI & 35.07$\pm$1.31 & 34.65$\pm$0.02 & {\bf\brown2.48$\pm$0.02} & 1.68$\pm$0.01 & {\bf\brown0.946$\pm$0.000} \\
HySure & 26.72$\pm$7.39 & 36.90$\pm$0.19 & 2.73$\pm$0.05 & 1.41$\pm$0.02 & 0.939$\pm$0.002 \\
LRSR & 55.78$\pm$0.69 & {\bf\red36.97$\pm$0.12} & 2.58$\pm$0.04 & {\bf\red1.36$\pm$0.02} & {\bf\red0.942$\pm$0.001} \\
NNM & {22.79$\pm$3.52} & {\bf\brown37.49$\pm$0.10} & {\bf\red2.51$\pm$0.03} & {\bf\brown1.34$\pm$0.01} & 0.940$\pm$0.001 \\
GLORIA & 44.38$\pm$5.46 & {\blue\bf38.23$\pm$0.01} & {\blue\bf2.30$\pm$0.00} & {\blue\bf1.22$\pm$0.00} & {\blue\bf0.949$\pm$0.000} \\
\hline
\end{tabular}%
}
\end{table}


\subsection{Synthetic Data Experiments}
\begin{figure}[H]
    \centering
    \begin{minipage}[c]{.28\linewidth}
        \centering
        \includegraphics[width=\linewidth]{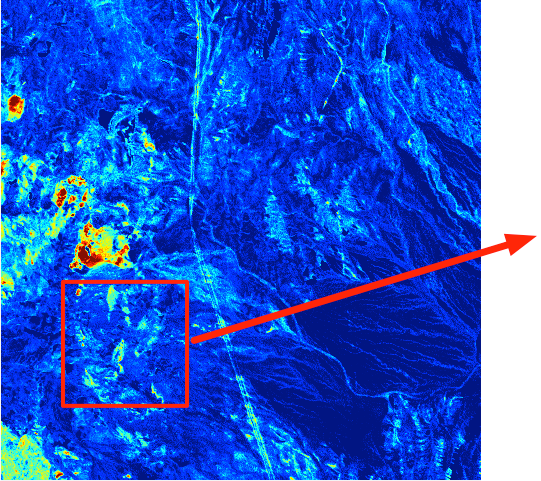}
        (a){\color{white}\qquad}
    \end{minipage}
    \begin{minipage}[c]{.25\linewidth}
        \centering
        \includegraphics[width=\linewidth]{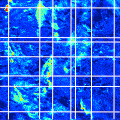}
        (b)
    \end{minipage}
    \caption{Synthetic SR image generation: (a) abundance map of one material (estimated from the Cuprite dataset); (b) segmentation of the abundance map.}
    \label{fig:syn_data}
\end{figure}

Second, we consider synthetic data experiments.
The way we prepare the experiment is similar to the one in \cite{simoes2015convex,veganzones2016hyperspectral}, with the difference that EV is also involved.
The procedure is described as follows.
We use the local-patch-wise, and EV-present, linear spectral mixture model \eqref{eq:i_am_tired} to generate the SR image.
The number of endmembers is $N= 5$.
The generations of the local endmember matrix $\bA_i$ and abundance matrix $\bS_i$ will be described shortly.
Each local patch is rectangular, but its horizontal and vertical lengths are, in each trial, random.
Fig.~\ref{fig:syn_data}(b) shows one such arrangement.
We obtain $P= 64$ such patches, and it is important to note that none of the algorithms, including GLORIA, has knowledge of such patch arrangement.
In GLORIA we will apply the same equi-spaced rectangular segmentation as before,
and there will be mismatches between the actual patches and the patches presumed by GLORIA.
Doing so is to provide a more realistic simulation as, in reality, it is impossible to exactly know how EV changes in space.

The endmember matrix $\bA_i$ is chosen as a collection of the spectral signatures of five materials, specifically, Actinolite, Albite, Muscovite, Olivine and Topaz.
To simulate the EV effect, for each patch and for each material we randomly pick one variation of that material from the U.S. geological survey (USGS) spectral library~\cite{kokaly2017usgs}.
The abundance matrix $\bS$ is chosen as the abundance maps extracted from a real HS image, namely, the AVIRIS Cuprite dataset;
the extraction is done by applying an HU method called SVMAX~\cite{chan2011simplex}.
In each trial, we randomly cropped a 120$\times$120 submap from AVIRIS Cuprite; see the illustration in Fig.~\ref{fig:syn_data}(a) where the submap is marked as a red rectangle.
The abundance maps are then extracted from that submap.

Some other simulation settings are as follows.
The spectral measurement response matrix corresponds to the spectral response of the Landsat 4 TM sensor.
The regularization parameter of GLORIA is $\gamma = 40/({\sf SNR}_{\rm M}  + {\sf SNR}_{\rm H})$.
We ran $100$ independent trials.
Table~\ref{table:syn_1} shows the results, where again the best three algorithms are marked in blue, brown and red.
As can be seen, GLORIA generally gives the best HSR recovery performance.
This suggests that GLORIA has the flexibility to accommodate the EV effect.
Also, GLORIA works better when the patch size is smaller, or when the number of patches $P$ increases.
Another observation is that at the lower SNR, i.e., $15$dB, GLORIA works considerably better than the state-of-art methods.
\begin{table}[!]
\centering
\caption{Average performance of the algorithms on the synthetic data.}
\label{table:syn_1}
\resizebox{.8\linewidth}{!}{%
\begin{tabular}{|c|c|c|c|c|c|} \hline
\multicolumn{6}{|c|}{${\sf SNR}_{\rm M}$/${\sf SNR}_{\rm H}$ - 15dB} \\\hline
\multicolumn{2}{|c|}{Method}  & PSNR & SAM & ERGAS & UIQI \\\hline
 \multicolumn{2}{|c|}{Ideal value}  & $\infty$ & 0 & 0 & 1 \\\hline
 \multicolumn{2}{|c|}{GSA}  & 10.98$\pm$1.27 & 19.29$\pm$4.71 & 9.49$\pm$1.83 & 0.216$\pm$0.055 \\
 \multicolumn{2}{|c|}{GLP}  & 14.76$\pm$0.32 & 11.09$\pm$0.18 & 5.90$\pm$0.17 & 0.273$\pm$0.048 \\
 \multicolumn{2}{|c|}{CNMF}  & 15.73$\pm$0.41 & 10.52$\pm$0.52 & 5.43$\pm$0.22 & 0.299$\pm$0.050 \\
 \multicolumn{2}{|c|}{FUMI}  & 17.63$\pm$0.16 & 7.69$\pm$0.42 & 4.32$\pm$0.17 & {\bf\red0.436$\pm$0.055} \\
 \multicolumn{2}{|c|}{HySure}  & 17.02$\pm$0.51 & 9.21$\pm$0.32 & 4.77$\pm$0.17 & 0.363$\pm$0.064 \\
 \multicolumn{2}{|c|}{LRSR}  & {\bf\red18.82$\pm$0.72} & {\bf\red6.98$\pm$0.54} & {\bf\red3.92$\pm$0.26} & {0.434$\pm$0.074} \\
 \multicolumn{2}{|c|}{NNM}  & 17.29$\pm$0.50 & 8.88$\pm$0.20 & 4.43$\pm$0.10 & 0.372$\pm$0.065 \\\hline
 \multirow{3}{*}{GLORIA} & $P=1$ & 18.46$\pm$0.23 & 7.90$\pm$0.10 & 4.00$\pm$0.06 & 0.432$\pm$0.060 \\
 & $P=9$ & {\bf\brown21.46$\pm$0.12} & {\bf\brown5.03$\pm$0.13} & {\bf\brown3.00$\pm$0.07} & {\bf\brown0.583$\pm$0.051} \\
 & $P=16$ & {\bf\blue21.76$\pm$0.14} & {\blue\bf4.61$\pm$0.10} & {\blue\bf2.83$\pm$0.06} & {\blue\bf0.598$\pm$0.053} \\\hline
 \multicolumn{6}{|c|}{${\sf SNR}_{\rm M}$/${\sf SNR}_{\rm H}$ - 25dB} \\\hline
 \multicolumn{2}{|c|}{GSA}  & 18.64$\pm$2.05 & 9.43$\pm$9.14 & 4.91$\pm$3.69 & 0.608$\pm$0.102 \\
 \multicolumn{2}{|c|}{GLP}  & 19.41$\pm$0.47 & 4.01$\pm$0.14 & 3.40$\pm$0.21 & 0.604$\pm$0.068 \\
 \multicolumn{2}{|c|}{CNMF}  & 24.71$\pm$0.35 & 3.89$\pm$0.20 & 1.97$\pm$0.10 & 0.729$\pm$0.048 \\
 \multicolumn{2}{|c|}{FUMI}  & 22.54$\pm$0.34 & 2.47$\pm$0.14 & 2.51$\pm$0.08 & 0.853$\pm$0.034 \\
 \multicolumn{2}{|c|}{HySure}  & {\bf\blue31.33$\pm$0.26} & {\bf\blue1.26$\pm$0.09} & {\bf\brown0.90$\pm$0.04} & {\bf\brown0.923$\pm$0.021} \\
 \multicolumn{2}{|c|}{LRSR}  & 30.63$\pm$0.22 & {\bf\red1.51$\pm$0.12} & 1.01$\pm$0.05 & {\bf\red0.914$\pm$0.021} \\
 \multicolumn{2}{|c|}{NNM}  & 27.30$\pm$0.44 & 2.78$\pm$0.05 & 1.42$\pm$0.03 & 0.820$\pm$0.045 \\\hline
 \multirow{3}{*}{GLORIA} & $P=1$ & 28.99$\pm$0.21 & 2.21$\pm$0.08 & 1.25$\pm$0.03 & 0.870$\pm$0.029 \\
 & $P=9$ & {\bf\red30.70$\pm$0.22} & 1.62$\pm$0.08 & {\bf\red1.00$\pm$0.03} & 0.909$\pm$0.023 \\
 & $P=16$ & {\bf\brown31.32$\pm$0.21} & {\bf\brown1.42$\pm$0.10} & {\bf\blue0.90$\pm$0.04} & {\bf\blue0.924$\pm$0.020} \\\hline
 \multicolumn{6}{|c|}{${\sf SNR}_{\rm M}$/${\sf SNR}_{\rm H}$ - 35dB} \\\hline
 \multicolumn{2}{|c|}{GSA}  & 21.31$\pm$2.55 & 7.85$\pm$11.01 & 4.38$\pm$4.35 & 0.796$\pm$0.109 \\
 \multicolumn{2}{|c|}{GLP}  & 21.07$\pm$0.53 & 1.68$\pm$0.09 & 2.79$\pm$0.19 & 0.782$\pm$0.070 \\
 \multicolumn{2}{|c|}{CNMF}  & 32.25$\pm$0.35 & 1.76$\pm$0.12 & 0.89$\pm$0.06 & 0.925$\pm$0.020 \\
 \multicolumn{2}{|c|}{FUMI}  & 23.45$\pm$0.30 & 1.47$\pm$0.11 & 2.32$\pm$0.09 & 0.949$\pm$0.010 \\
 \multicolumn{2}{|c|}{HySure}  & {\bf\red35.93$\pm$0.42} & {\bf\red1.02$\pm$0.10} & {\bf\red0.55$\pm$0.04} & {\blue\bf0.977$\pm$0.006} \\
 \multicolumn{2}{|c|}{LRSR}  & 34.89$\pm$0.40 & 1.03$\pm$0.11 & 0.63$\pm$0.04 & {\bf\red0.971$\pm$0.007} \\
 \multicolumn{2}{|c|}{NNM}  & 34.87$\pm$0.32 & 1.19$\pm$0.02 & 0.59$\pm$0.01 & 0.959$\pm$0.014 \\\hline
 \multirow{3}{*}{GLORIA} & $P=1$ & 35.48$\pm$0.51 & 1.09$\pm$0.08 & 0.58$\pm$0.05 & 0.966$\pm$0.012 \\
 & $P=9$ & {\bf\brown36.18$\pm$0.22} & {\bf\brown1.01$\pm$0.03} & {\bf\blue0.52$\pm$0.01} & 0.970$\pm$0.011 \\
 & $P=16$ & {\blue\bf36.73$\pm$0.49} & {\blue\bf0.96$\pm$0.09} & {\brown\bf0.53$\pm$0.04} & {\bf\brown0.974$\pm$0.008} \\\hline
\end{tabular}%
}
\end{table}

Previously we mentioned that GLORIA, an inexact MM scheme using accelerated projected gradient, runs very fast.
To give the reader some idea,
we conduct the following runtime test.
We benchmark GLORIA against the exact MM scheme and the inexact MM scheme via the nominal projected gradient (PG).
The exact MM scheme solves each MM iteration exactly via the accelerated PG method (see Section~\ref{sec:APG}).
The inexact MM scheme via the nominal PG removes the extrapolation by setting $\alpha_k= 0$ in the inexact MM iteration \eqref{eq:iMM_iter}.
The previously described synthetic experiment is used to test the three MM schemes,
and we consider SNR= $25$dB and $P= 16$.
The runtime results, shown in Table~\ref{tab:runtimes}, indicate significant runtime advantages of GLORIA over the other two candidates.
\begin{table}[!]
    \centering
    \caption{Runtime performance of various MM schemes.}
    \label{tab:runtimes}
    \begin{tabular}{|c|c|c|c|}
        \hline
         & Exact MM  & Inexact MM via nominal PG & Inexact MM via APG (GLORIA) \\\hline
        Time (sec.) & 214.16$\pm$114.83 & 102.48$\pm$2.44 & 24.25$\pm$0.35 \\
        \hline
    \end{tabular}
\end{table}
\clearpage
\subsection{Real Data Experiments}
\begin{figure}[H]
\hfil
    \begin{minipage}[c]{.25\linewidth}
        \centering
        \includegraphics[width=\linewidth]{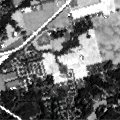}
    \end{minipage}\qquad
    \begin{minipage}[c]{.25\linewidth}
        \centering
        \includegraphics[width=\linewidth]{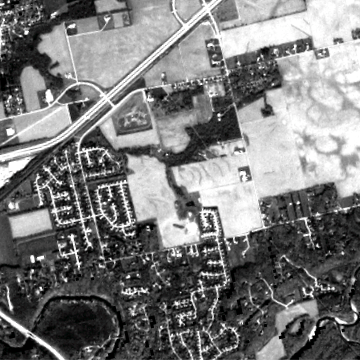}
    \end{minipage}
    \caption{The tested HS-MS image pair: left) the 5th band of the HS image, right) the 1st band of the MS image}
    \label{fig:pair}
\hfil
\end{figure}

Finally, we test the algorithms on real data.
The dataset is the one used in~\cite{yang2018hyperspectral}.
The HS image was acquired by the Hyperion HS sensor.
It covers a spectral range from 400nm to 2,500nm, and has 89 spectral bands after removing 131 noisy bands.
The MS image was captured by the MS sensor mounted on the Sentinel-2A satellite.
It has 13 bands, and we adopt 4 bands whose central wavelengths are 490nm, 560nm, 665nm and 842nm, resp.
Readers are referred to \cite{yang2018hyperspectral} for further details.
After pre-processing such as co-registration and cropping, we obtain the HS-MS image pair.
The image pair is illustrated in Fig.~\ref{fig:pair}.
The setting is $(M,\Mm,L,\Lh,N)=(89,4,360^2,120^2,30)$.
We employ the algorithm in \cite{simoes2015convex} to estimate the spectral and spatial measurement responses $\bF$ and $\bG$.
Empirically we found that, for the tested HS-MS pair, this algorithm happens to yield a poor estimate of $\bG$.
To mend this issue, we consider a second-stage estimation:
we fix the blurring kernel to be a Gaussian kernel and estimate its standard deviation $\sigma^2$ by the nonlinear least-squares $\min_{\sigma^2>0}\|\by_{\rm M}^1\circ g(\sigma^2)-\bm f^1\bY_{\rm H}\|^2$,
where $g(\sigma^2)$ denotes the Gaussian kernel with variance $\sigma^2$, $\circ$ corresponds to the 2-D convolution operation, and $\bym^1$ and $\bm f^1$ are the first row of $\bYm$ and the estimated $\bF$, resp.

\begin{figure}[H]
\begin{minipage}[c]{.98\textwidth}
    \begin{minipage}[c]{.244\textwidth}
        \centering
        \includegraphics[width=\linewidth]{figs/band5/realdata_0.png}
        (a) HS image
    \end{minipage}
    \begin{minipage}[c]{.244\textwidth}
        \centering
        \includegraphics[width=\linewidth]{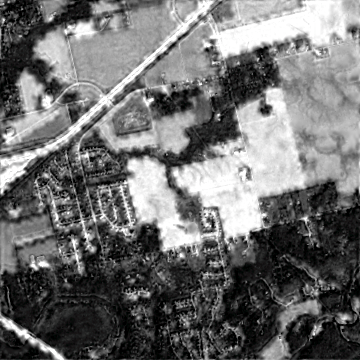}
        (b) GLP
    \end{minipage}
    \begin{minipage}[c]{.244\textwidth}
        \centering
        \includegraphics[width=\linewidth]{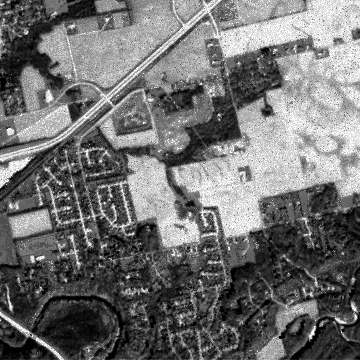}
        (c) CNMF
    \end{minipage}
    \begin{minipage}[c]{.244\textwidth}
        \centering
        \includegraphics[width=\linewidth]{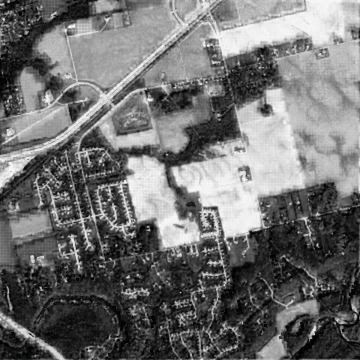}
        (d) FUMI
    \end{minipage}
    \begin{minipage}[c]{.244\textwidth}
        \centering
        \includegraphics[width=\linewidth]{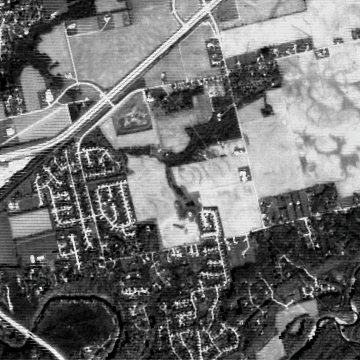}
        (e) HySure
    \end{minipage}
    \begin{minipage}[c]{.244\textwidth}
        \centering
        \includegraphics[width=\linewidth]{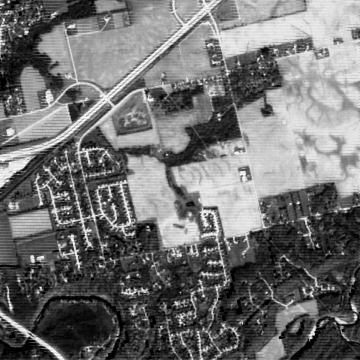}
        (f) LRSR
    \end{minipage}
    \begin{minipage}[c]{.244\textwidth}
        \centering
        \includegraphics[width=\linewidth]{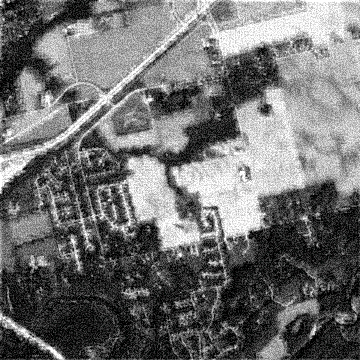}
        (g) NNM
    \end{minipage}
    \centering
    \begin{minipage}[c]{.244\textwidth}
        \centering
        \includegraphics[width=\linewidth]{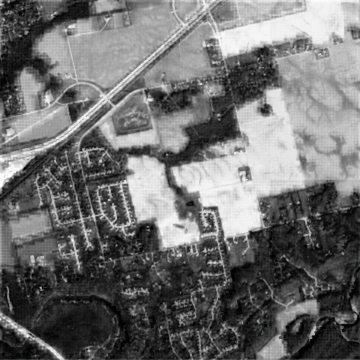}
        (h) GLORIA
    \end{minipage}
\end{minipage}
\caption{The 5th bands of the HS image and recovered images}
\label{fig:5th}
\begin{minipage}[c]{.98\textwidth}
    \begin{minipage}[c]{.244\textwidth}
        \centering
        \includegraphics[width=\linewidth]{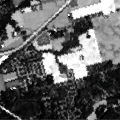}
        (a) HS image
    \end{minipage}
    \begin{minipage}[c]{.244\textwidth}
        \centering
        \includegraphics[width=\linewidth]{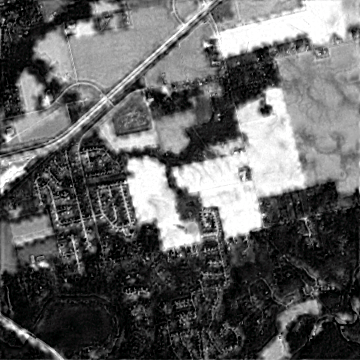}
        (b) GLP
    \end{minipage}
    \begin{minipage}[c]{.244\textwidth}
        \centering
        \includegraphics[width=\linewidth]{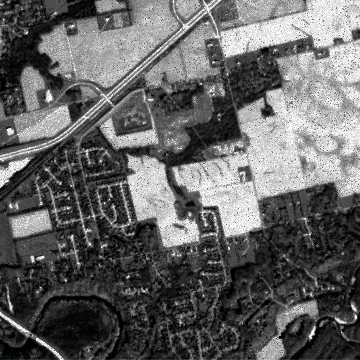}
        (c) CNMF
    \end{minipage}
    \begin{minipage}[c]{.244\textwidth}
        \centering
        \includegraphics[width=\linewidth]{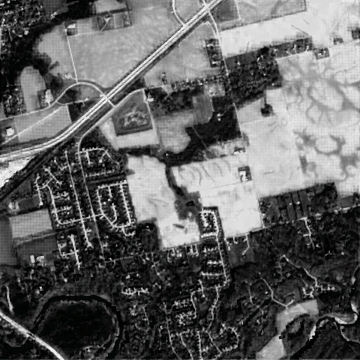}
        (d) FUMI
    \end{minipage}
    \begin{minipage}[c]{.244\textwidth}
        \centering
        \includegraphics[width=\linewidth]{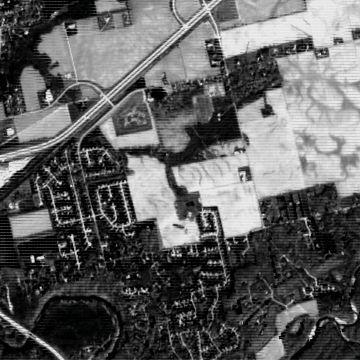}
        (e) HySure
    \end{minipage}
    \begin{minipage}[c]{.244\textwidth}
        \centering
        \includegraphics[width=\linewidth]{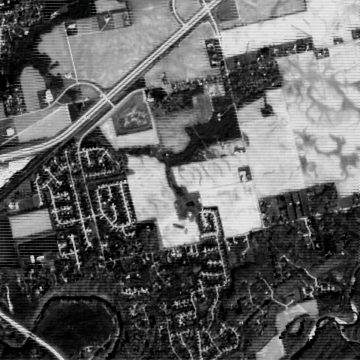}
        (f) LRSR
    \end{minipage}
    \begin{minipage}[c]{.244\textwidth}
        \centering
        \includegraphics[width=\linewidth]{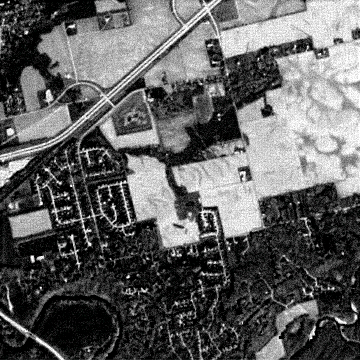}
        (g) NNM
    \end{minipage}
    \centering
    \begin{minipage}[c]{.244\textwidth}
        \centering
        \includegraphics[width=\linewidth]{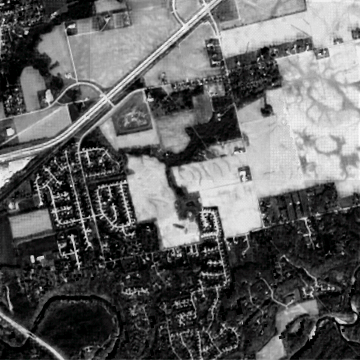}
        (h) GLORIA
    \end{minipage}
\end{minipage}
\caption{The 20th bands of the HS image and recovered images}
\label{fig:20th}
\end{figure}

We follow the same simulation settings as those in semi-real experiments, except that we use $\gamma=10$ for GLORIA.
Figs.~\ref{fig:5th}-\ref{fig:20th} illustrate the 5th and 20th bands of the original MS image and recovered images.
From the figures, we note that GLP does not work well compared to the other algorithms.
Moreover, if we zoom in the recovered images, we can see that the images recovered by HySure and LRSR have strip noise, while the images recovered by CNMF and NNM have pepper noise.
In comparison, the images recovered by FUMI and GLORIA appear smoother.

\section{Conclusion}

In this paper we explored the route of low-rank matrix estimation for HSR.
By positing a low-rank spectral-spatial data structure, both globally and locally,
we built an algorithmic solution, called GLORIA, that exploits such structure for HSR.
Our extensive numerical studies, which include semi-real data experiments on five different datasets, one synthetic experiment and one real experiment, show that exploiting global-local low-rank structure not only is a working idea but also provides satisfactory reconstruction results.
Our global-local low-rank  exploitation is made possible by customizing an efficient first-order strategy in large-scale structured optimization.
We close this paper by naming a future direction.
It would be interesting to study how low-rank matrix estimation would be useful in other hyperspectral problems such as multisource and multitemporal data fusion.

\vspace*{-.75em}
\section*{Appendix A}
\label{proof:lip}

By definition, a constant $L$ is said to be a Lipschitz constant of $\nabla g_k$ on $\setX$ if
$\| \nabla g_k(\bX)-\nabla g_k(\bY)\|_F \leq L \| \bX - \bY \|_F$ for any $\bX, \bY \in \setX$.
From  \eqref{eq:nabla_gk} we have
\begin{subequations}
	\begin{align}
	&\|\nabla g_k(\bX)-\nabla g_k(\bY)\|_F\\
	\leq&\|(\bF^T\bF+p\gamma_0\bW_0^k)(\bX-\bY)\|_F+\|(\bX-\bY)\bG\bG^T\|_F\nonumber\\
    &\qquad\qquad\qquad\qquad+p\sum_{i=1}^P\|\gamma_i\bW_i^k(\bX_i-\bY_i)\|_F
	\label{eq:app_eq1} \\
	\leq&\left(\lammax(\bF^T\bF+p\gamma_0\bW_0^k)+\lammax(\bG\bG^T)\right.\nonumber\\
    &\qquad\qquad\left.+p\max_{i=1,\ldots,P}\gamma_i \lammax (\bW_i^k)\right)\|\bX-\bY\|_F.
	\label{eq:app_eq2}
	\end{align}
\end{subequations}
for any $\bX,\bY\in\Rbb^{M\times L}$,
where \eqref{eq:app_eq1} is due to the triangle inequality;
\eqref{eq:app_eq2} is due to i) the inequality $\| \bA \bB \|_F \leq \| \bA \|_2 \| \bB \|_F$ in which $\| \bA \|_2$ denotes the spectral norm of $\bA$, and ii) the identity $\| \bA \|_2 = \lammax(\bA)$ for positive semidefinite $\bA$.
The proof of Fact~\ref{fac:lip} is complete.


\bibliographystyle{IEEEtran}
\bibliography{refs}

\end{document}